\documentclass[amsmath,amssymb,aps,pre,superscriptaddress,citeautoscript,floatfix]{revtex4-1}
\usepackage{graphicx}
\usepackage[bookmarks=false]{hyperref}
\usepackage{xcolor}

\def\bnabla{{\boldsymbol{\nabla}}}
\newcommand {\vct}[1] {\mathbf {#1}}

\begin{document}
\title{Ordering of adsorbed rigid rods mediated by the Boussinesq interaction on a soft substrate}

\author{Sunita Kumari}
\affiliation{School of Physical Sciences, University of Chinese Academy of Sciences, Beijing 100049, China}
\author{Fangfu Ye}
\affiliation{CAS Key Laboratory of Soft Matter Physics, Institute of Physics, Chinese Academy of Sciences, Beijing 100190, China}
\affiliation{Wenzhou Institute of the University of Chinese Academy of Sciences, \\
Wenzhou, Zhejiang 325000, China}
\author{Rudolf Podgornik}
\email{podgornikrudolf@ucas.ac.cn}
\affiliation{School of Physical Sciences and Kavli Institute for Theoretical Sciences, University of Chinese Academy of Sciences, Beijing 100049, China}
\affiliation{CAS Key Laboratory of Soft Matter Physics, Institute of Physics, Chinese Academy of Sciences, Beijing 100190, China}
\affiliation{Department of Physics, Faculty of Mathematics and Physics, University of Ljubljana, 1000 Ljubljana, Slovenia}


\begin{abstract}
Orientational ordering driven by mechanical distortion of soft substrates plays a major role in material transformation processes such as elastocapillarity and surface anchoring. We present a theoretical model of the orientational response of anisotropic rods deposited onto a surface of a soft, elastic substrate of finite thickness. We show that anisotropic rods exhibit a continuous {\sl isotropic-nematic phase transition}, driven by orientational interactions between surface deposited rods. This interaction is mediated by the deformation of the underlying elastic substrate, and is quantified by the Boussinesq solution adapted to the case of slender, surface deposited rods. From the microscopic rod-rod interactions we derive the appropriate Maier-Saupe mean-field description, which includes the Boussinesq elastic free-energy contribution due to the substrate elasticity, derive the conditions for the existence of a continuous orientational ordering transition  and discuss the implication of results in the soft (bio) systems context.
\end{abstract}
\maketitle

\section{\label{sec:1}Introduction}
Living organisms interact sensitively with their  environments in particular with numerous surfaces and substrates permeating their natural {\sl milieu}. The surface compliance triggers varied responses that play key roles in regulating their behaviors and fates. For example, the novel coronavirus (SARS-CoV2) deposited on various surfaces can proliferate for varied amounts of time, depending on the nature of the substrate~\cite{neeltje2020}. Spinal cord neurons extend more primary dendrites and shorter axons on stiffer gels~\cite{jiang2006} and astroglia are less adherent to the softer hydrogels compared to hard gels~\cite{georges2006}. Rod-like fd virus particles form three distinct interaction regimes (linear chains, collapsed globules and chain-like aggregation) on lipid membranes  depending  on  the  adhesion strength~\cite{petrova2017}. Interestingly, stationary cells plated on an elastic substrate which is cyclically stretched re-orientate away from the stretching direction~\cite{dar86, grood00}, and locomotive cells on a strained elastic substrate re-orient in the strain direction~\cite{cmlo00}. 

As design and technology entails progressively more sophisticated manipulation of materials, the role of surfaces is becoming more prominent. In this context, a soft elastic substrate is of particular interest because of its flexibility and deformability  that can mediate long-range elastic interactions between substrate-embedded particles~\cite{dan1993,rudnick1999, weikl2003, dean2012}, often giving rise to different structural motifs, such as an analog of elastic ``Cheerios effect" in which spherical~\cite{aditi2013} and cylindrical~\cite{aditi2015} particles are suspended in an elastic gel. The long-range elastic interactions between particles deposited on a soft gel surface originate from the local surface deformation of the substrate, created either by their weight, by the adsorption energy,  or can be imposed from external constraints/fields such as, e.g., by external electric or magnetic fields or field gradients, and as a consequence lead to long-range attractive interaction ~\cite{aditi2013, aditi2015,menzel2017,puljiz2017}. Similar interactions are found for liquid droplets on elastomeric interfaces that give rise to an inverted Cheerios effect~\cite{stefan2016}. The effect of the surface elasticity has been studied also in the context of wetting transition and adsorption phenomena~\cite{bernardino2012, bruno2020, napio2020}. Elastic substrates introduce complex liquid film dynamics with surface topography as the substrate shape changes with adsorption and in turn alters its wetting properties, modifying the adsorption properties of the vicinal liquid~\cite{bernardino2012}. Combined effects of surface elasticity and wetting have been demonstrated also in the context of ultrahydrophobic surfaces~\cite{otten2004,mock2005, thomas2016} and fast condensation on soft surfaces~\cite{sokuler2010}. Elastic interactions in conjunction with defects in liquid crystals control the coordinates of embedded cylinders and tune the directional interactions~\cite{liu2015} so that the trapped cylinders orient along the director axis on thin nematic films and pairs of cylinders form chains by capillarity, which forces them to align along the director by an elastic torque~\cite{liu2015}.

The properties of thin membranes containing anisotropic inclusions were examined by considering the membrane as a continuous medium~\cite{podgornik1993}. In the linear response approximation the membranes were shown to be weakly perturbed by inclusions of particles and to mediate long-range elastic interactions between the inclusions~\cite{dan1993}. Using an analytical method, a wrapping - unwrapping transition of an elastic filament around cylinders has been studied in detail and a change of entropy due to wrapping of the filament around the adsorbing cylinders as they move closer together was identified as an additional source of interactions between them~\cite{dean2012}. Recent computer simulation work shows that the rigid rod-like particles on an elastic substrate display  membrane-driven self-organization controlled by the tension and curvature of the underlying membrane~\cite{ghosh2016}. An effective field theory was derived for the analytical computation of the  membrane-mediated interactions, reducing the rigid particles to points and computing the interaction free energy as an asymptotic expansion in inverse separations ~\cite{yolcu2012}. Within the framework of the free energy formulation, the intricate shapes of hydrogel menisci due to the indentation of rigid particle has also been analyzed and shown to result from a competition between surface tension, elasticity and hydrostatic pressure inside the gel, which drive the mutual interaction between the particles ~\cite{pandey2018}. In a very recent work, a mean-field theory (MFT) based on the minimization of a free energy functional was implemented to explain the wetting transition of elastic substrates~\cite{napio2020}. The substrate was described in terms of the linear theory of elasticity, and the fluid contribution was based on the van der Waals theory.  Substrate elasticity was shown to imply long-range attractive interaction between fluid particles, which moreover induced a decrease of the critical wetting temperature as compared to the case of wetting on an inert substrate, or precluded critical wetting of the elastic substrate altogether~\cite{napio2020}.

Inclusion problem in an infinite elastic matrix has a venerable history~\cite{eshelby,thein1979}.  Phan-Thien calculated the total force exerted on a fibre in a homogeneous displacement field in the limit of low concentrations of the fibres  so that interaction effects between different fibres can be neglected~\cite{thein1979}, while Eshelby considered external elastic field  outside an ellipsoidal inclusion or a general elastic inhomogeneity~\cite{eshelby}. However, in both cases and contrary to the work described in this contribution, the analysis is restricted to an infinite elastic medium. Indeed, in many applications the size of the host elastic matrix, in which an inclusion is embedded, is never infinite and may be not large enough to be approximated as such in comparison with the inclusion size. This limitation becomes obvious, when the size effect and boundary effects of an inhomogeneity become prominent issues. In order to include the  finite-domain effects, a bounded homogeneous elastic medium with volume inclusions was considered in Ref. ~\cite{mura1984}. An exact closed-form solution was derived for the Dirichlet–Eshelby or Neumann–Eshelby tensors depending on which of the displacement-free or traction-free boundary conditions is imposed on the outer boundary of the  elastic matrix~\cite{li2005,wang2005}. These works were further extended in Refs~\cite{gao2010,gao2011} from the classical linear elasticity to a strain gradient elasticity theory. The periodic Eshelby inclusion problems was solved in Ref.  ~\cite{liu2010}, while the problem of inclusions axisymmetric with respect to enclosing spherical domain was investigated in Ref. ~\cite{mejak}

Despite significant progress in recent years on the understanding of soft surfaces, the study of orientational interactions, mediated by the underlying elastic matrix, on the properties of particles deposited on deformable substrates is comparatively scant, even if the perturbed orientational degrees of freedom certainly play an important role. In what follows, we embark on a comprehensive analysis of the elastic substrate-mediated, ordered, nematic-like phases of surface deposited rods. To the best of our knowledge, no work has been performed yet to discuss an isotropic-nematic phase transition of rods deposited onto an elastic substrate. Studying such systems can provide insights into the influence of confined substrate deposition on physical properties of soft matter.

We consider in a comprehensive manner a system of anisotropic rods deposited onto a surface of a soft, elastic substrate of finite thickness. The substrate deforms under the loading of rods and this deformation mediates an orientational interaction between them. We apply the Boussinesq theory ~\cite{joseph85}, modified due to the presence of a finite thickness of the substrate, to calculate the deformation and the consequent elastic displacements within and on the surface of the substrate. The original Boussinesq theory describes the displacement equilibrium inside an elastic half-space for point-like forces normal to the surface~\cite{love29}. We  generalize this theory to account for a finite thickness of the elastic medium as well as the rod-like shapes of the inclusions. The finite thickness effects are analysed straighforwardly within the {\sl Galerkin representation} of the deformation vector, with the appropriate boundary conditions at the (upper) surface with a fixed normal force and the restrained (bottom) surface of a finite-thickness elastic slab. The rod shape and the ensuing orientational effects are taken into account by integrating the {\sl Green function} of the finite slab Boussinesq problem over the length of the rod, in several respects similar to the methodology of Phan-Thien~\cite{thein1979} except that the Boussinesq Green function is substituted for the infinite volume "Kelvinlets". The corresponding rod-rod long-range directional orientation pair-wise interactions are subsequently expanded in terms of the orientations of the rods and a canonical thermal ensemble of rods is then analysed within a collective field-theoretical statistical mechanical description. A mean-field approximation for the field-theoretical partition function is pursued explicitly, leading straightforwardly to the Maier-Saupe framework for the orientational degrees of freedom and to the van der Waals condensation for the positional degrees of freedom. Only the former is investigated in detail. While the elastic surface-deposited rod problem is not strictly 2 dimensional (2D) - the elastic interactions mediated by the finite thickness slab are obviously 3D - the Maier-Saupe equation looks like the one for a 2D system, implying a continuous orientational phase transition with a quasi-long-range order (that is, the correlations in orientational order decay algebraically) just like in a real 2D nematic~\cite{Frenkel2000, frenkel,Lago2003}.

Our work covers comprehensively all the aspects of the problem, from the derivation of the microscopic pair potential and then all the way to statistical properties of the thermal ensemble. The paper itself is organized as follows: To demonstrate the underpinnings of the theory and for an intuitive illustration of the methodology, we first consider a single point particle interaction with the elastic substrate in Sec.~\ref{sec:2}. The phenomenology of the elastic deformation is described briefly and we derive explicitly the Green functions associated with an incompressible linear elastic solid by solving the pertaining Navier equation within the standard Galerkin {\sl Ansatz}. In the next Sec.~\eqref{sec:3}, the pairwise interactions between two rod-like particles, obtained by analogy with electrostatics ~\cite{deutsch}, which on the other hand is just an extension of the Phan-Thien approach ~\cite{thein1979} for an infinite elastic medium, are considered together with the many-particle generalization. The orientational dependence of the interactions in the limit of large separations between the rods is then derived within a second order ``multipole" expansion. In Sec.~\eqref{sec:4}, we subsequently formulate a field-theoretical representation of the partition function for a canonical ensemble of $N$ rods deposited onto the elastic surface. The partition function is calculated explicitly within a mean-field (saddle-point) approximation for the density and the orientational order parameters. We delineate some conclusions and discuss the limitations of our approach in the last Sec.~\eqref{sec:5}. Finally, in the Appendices we present all the pertinent computational details.
 
\section{\label{sec:2}The Boussinesq problem with two point-like sources}

\begin{figure*}
\includegraphics[width=16cm]{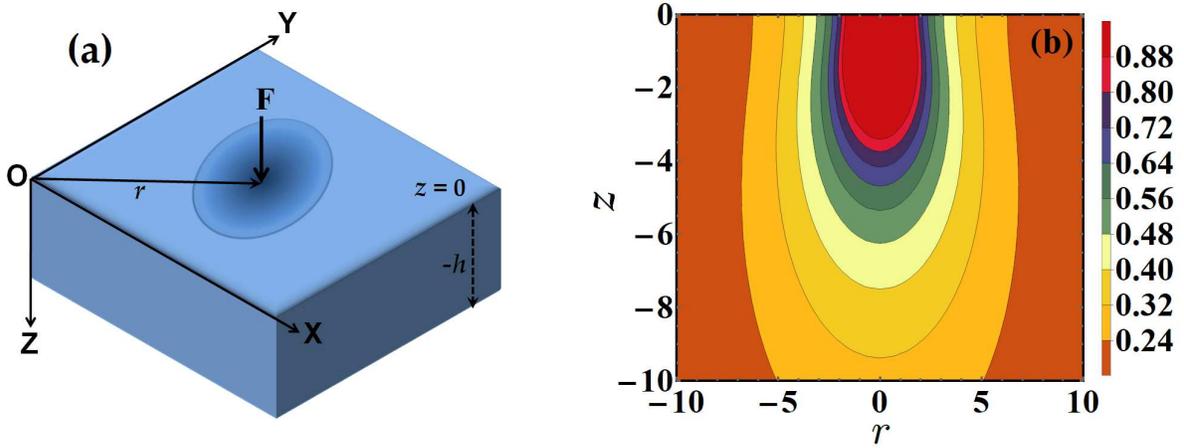}
\caption{The Boussinesq problem for a finite thickness elastic substrate. (a) The deformed surface profile (schematic) of a substrate for a point-like source $\vct{F}$ at $ r(x,y)$ acting at the upper $z=0$ plane of a substrate. (b) Contour plot of the Boussinesq iso-$u_z$ ($u_z(r, z)$) surfaces from Eq.~\eqref{eq:38} for $\sigma =0.5$. We plot $u_z(r, z)$, $z$ and $r$ in the units of $\sqrt{(\rho_s (1+\sigma) P)/2\pi E}$.}
\label{point_force}
\end{figure*}

We start with the the problem of elastic deformations and stresses arising from a single, rigid point-like particle deposited on the surface of an elastic substrate of finite thickness, see Fig.~\eqref{point_force}. The later condition is not considered in the standard {\sl Boussinesq problem}, which is formulated for a semi-infinite substrate~\cite{joseph85}. The Boussinesq theory is based on elastic displacement potential function for a point-like normal force, acting at the surface, which satisfies the Navier equation and which can be used to find the displacement components~\cite{love29}. A related problem of a tangential force acting at the surface is considered in the {\sl Cerruti problem}~\cite{Cerruti}.

Throughout this paper, we consider an isotropic and homogeneous elastic substrate of a finite thickness $h$. We assume a point-like rigid particle centered at $(x, y, z=0) = ({\bf r}, z=0)$, with ${\bf r} = (x,y)$ at the upper free surface of the substrate, chosen to be at the $z = 0$ plane. The $z$ axis points downward (negative) so that points within the substrate have $z < 0$. The point-like force $\vct{F}=(0,0,f)$ imposed by this particle creates a stress distribution in the substrate and as a result the substrate is deformed. The exact nature of this applied force is immaterial at this point. It can be either externally imposed, can be the weight of the inclusion, or can be a result of the adhesion energy between the particle and the surface, see Sec.~\eqref{sec:5} for discussion. 

The elastic stresses are obviously distributed with cylindrical symmetry and we formulate the equations for the displacement field at any arbitrary point in the substrate due to the applied force. The surface density of the force in the direction normal to the surface is given by $F_z = P\delta^2({\bf r}-{\bf r}')$, where $\delta^2({\bf r})$ is the 2D Dirac delta function so that {\sl in extenso}
\begin{equation}
    \vct{F}=(0,0,F_z({\bf r}, z)) = (0, 0, P ~\delta(z) \delta^2({\bf r}-{\bf r}')),
    \label{gfdhjksa0}
\end{equation}
with $P$ its magnitude. In the linear regime, the elastic displacement fields must satisfy the Navier equation~\cite{LL},
\begin{eqnarray}
\nabla ^{2}\vct{u}({\bf r},z) + \frac{1}{1-2\sigma}\nabla \nabla \cdot \vct{u}({\bf r},z)  = - \frac{2\rho_s (1+\sigma)}{E} \vct{F}({\bf r}, z), 
\label{navier}
\end{eqnarray}
where $\rho_s$ is the density of the substrate material, $\sigma$ is the Poisson ratio related to the compressibility of the substrate and $E$ is its elastic modulus. For isotropic materials, the Poisson ratio must satisfy $-1 \leq \sigma \leq .5$. In general, Eq.~\eqref{navier} is difficult to solve because of the coupling between various components of the displacement vector. However, it can be solved straightforwardly by introducing the {\sl Galerkin vector}  representation method, which transforms the second order Navier differential equation into a fourth-order Galerkin equation (for detailed derivation see Appendix~\eqref{appendix:a}) by introducing $\mathbf{g}$~\cite{gal36},
\begin{eqnarray}
\mathbf{u} = \nabla^2\mathbf{g} - \frac{1}{2(1 - \sigma)}\nabla (\nabla \cdot \mathbf{g}).
\label{solveg}
\end{eqnarray}

Because of the cylindrical symmetry of the problem this equation can be simplified further by introducing the {\sl Love potential function}~\cite{love29} as,
\begin{eqnarray}
\mathbf{g} = (0,0,Z({\bf r}, z)), 
\label{g=Z}
\end{eqnarray}
in terms of which we remain with a scalar inhomogeneous biharmonic equation for the Love potential,
\begin{eqnarray}
\nabla^{2} \nabla^{2}Z({\bf r},z) = \nabla^{4}Z({\bf r},z) = -\frac{2\rho_s (1 + \sigma)}{E} F_z({\bf r}, z).
\label{eq:39}
\end{eqnarray}
The connection between the displacement vector $\bf u$, the stress tensor $p_{ij}$ and the Love potential can be derived straightforwardly (for detailed derivation see Appendix \eqref{appendix:a}). Since the external force is located at the upper surface of the substrate, it can be subsummed into a boundary condition, leaving a homogeneous biharmonic equation (Eq.~\eqref{eq:39} with vanishing r.h.s.).  

The boundary conditions for the solution of the homogeneous Love equation, Eq.~\eqref{eq:39}, reduce to the pressure tensor components on the upper surface of the substrate, given by the external point-like surface force, Eq.~\eqref{gfdhjksa0}, and vanishing of the elastic displacement on the lower surface. Thus, from Eq.~\eqref{p_xy} - \eqref{p_zz} the boundary conditions read
\begin{subequations}
\begin{equation}
p_{xz} = p_{yz} = 0,~~~~ p_{zz} = P ~\delta^2({\bf{r}}), 
~~~~{\rm at} ~z = 0,
\label{pxpy_01}
\end{equation}
and 
\begin{equation}
    u_x=u_y=u_z = 0,~~~~{\rm at} ~z = -h.
    \label{pz_0}
\end{equation}
\end{subequations}
Clearly the biharmonic equation for the Love potential, Eq.~\eqref{eq:39}, bears a lot of similarity to the Poisson equation of classical electrostatics and can be solved by the same devices, specifically by the introduction to the elastic dyadic Green function, whose $zz$ component equals the $z$-component of the strain vector for a delta function source.

\begin{figure}[t!]
\includegraphics[width=17cm]{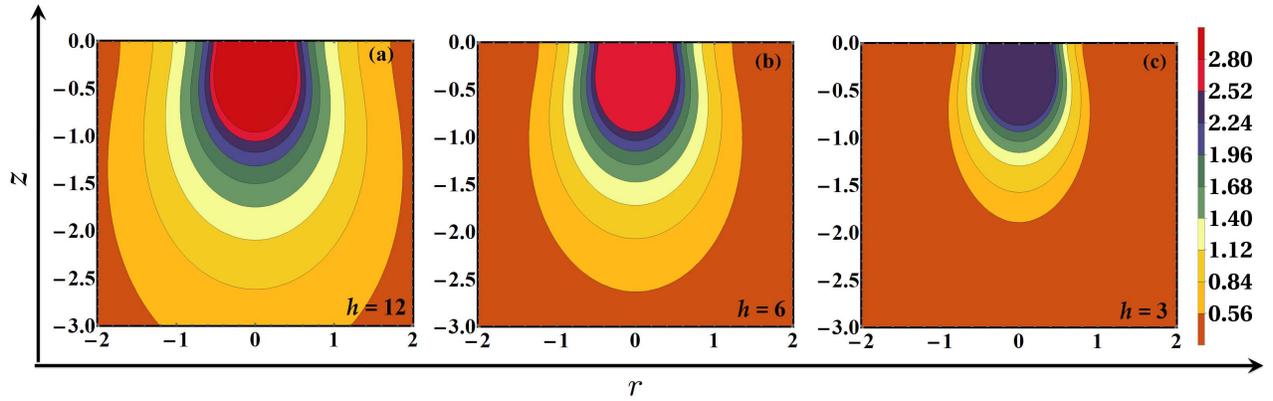}
\caption{Contour plots of $u_z(r, z)$ for $\sigma = 0.5$ and different values of the substrate thickness $h = 12, 6, 3$ (a, b, c). Because of the boundary condition at $z=h$, stipulating that all the components of the deformation vector should vanish, ${\bf u}(Q, z = -h) = 0$, the deformation profile is contracted in the $z$-direction for smaller $h$. This contraction is the most important feature of the finite layer Boussinesq problem. We plot $u_z(r, z)$, $z$ and $r$ in the units of $\sqrt{(\rho_s (1+\sigma) P)/2\pi E}$.}
\label{uzprofile}
\end{figure}

Consistent with the cylindrical symmetry of the problem one can introduce the Fourier-Bessel transform for all the variables, {\sl i.e.}, the displacement components as well as $Z(Q,z)$ 
\begin{eqnarray}
Z(r,z) = \frac{1}{2\pi}\int_{0}^{\infty}QdQJ_{0}(Qr)Z(Q,z),
\label{ft_stress_function}
\end{eqnarray}
with $r = \vert {\bf r}\vert$, in terms of which the homogeneous Love equation is reduced to 
\begin{eqnarray}
\left(\frac{\partial ^2 }{\partial z^2} - Q^2\right)^{2}Z(Q,z) = 0,
\label{def_z}
\end{eqnarray}
and has a general solution:
\begin{eqnarray}
Z(Q,z) = (A + BQz)e^{-Qz} + (C + DQz)e^{Qz},
\label{four_coefficients}
\end{eqnarray}
where the coefficients can be obtained from the boundary conditions Eqs.~\eqref{pxpy_01} and \eqref{pz_0}. The displacement field then follows as, see Eqs.~\eqref{displacement_xy} and \eqref{displacement_z}, 
\begin{subequations}
\begin{eqnarray}
u_{x,y}(r, z) &=& - \frac{1}{2(1 - \sigma)}\frac{\partial}{\partial x,y}~\frac{1}{2\pi}\int_{0}^{\infty}QdQJ_{0}(Q r) \frac{\partial Z(Q,z) }{\partial z} 
\label{displacement_xy1}
\end{eqnarray}
\begin{eqnarray}
u_z(r, z) = \frac{1}{2\pi}\int_{0}^{\infty}QdQJ_{0}(Q r)\left(Q^2 Z(Q,z) - \frac{1 - 2\sigma}{2(1 - \sigma)}\frac{\partial^{2}Z(Q,z)}{\partial z^{2}}\right).
\label{FBGreens1}
\end{eqnarray}
\end{subequations}
The corresponding contours of the $u_z(r, z)$ profiles, obtained from Eqs.~\eqref{FBGreens1}, are displayed in Fig.~\eqref{uzprofile}. Clearly the effect of the boundary condition at the lower $z=-h$ surface of the substrate, is to ``compress" the contours in the $z$-direction. The extent of the deformation due to the surface source is thus cut down mostly to the vicinity of the top layer.

\begin{figure*}[t!]
\includegraphics[width=17cm]{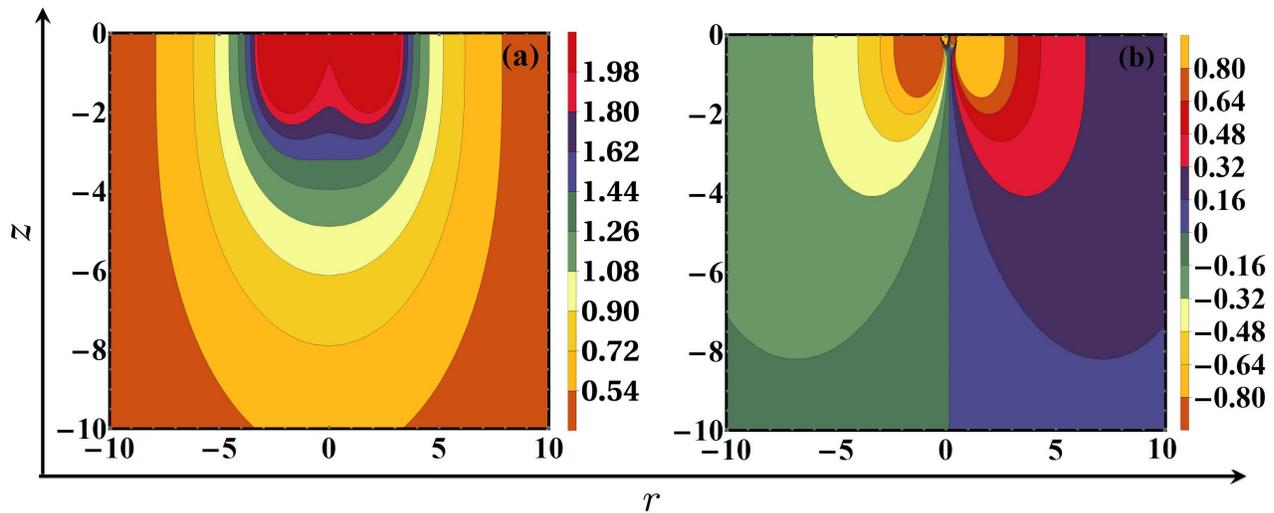}
\caption{(a) The contours of $u_z(r, z)$ for two point sources with $\sigma = 0.01$. (b) The contours of $u_{x,y}(r, z)$ for two point sources. Both are obtained from a linear superposition of two point-like solutions of the Boussinesq problem from Eq.~\eqref{eq:38}. The combined deformation for the two sources ${\bf u}(r, z)$, mostly in the region between the sources, leads to a more negative total energy and thus to an effective elastic attraction between the sources. We plot $u_z(r, z)$, $u_{x,y}(r, z)$, $z$ and $r$ in the units of $\sqrt{(\rho_s (1+\sigma) P)/2\pi E}$.}
\label{two_points}
\end{figure*}

Building on the obvious analogy with electrostatics - a valid analogy since both theories are linear - but taking into account that the fields and sources are vectorial in the Boussinesq problem, one can obtain the resulting deformation vector for two point-like vector sources $P_1, P_2$ acting at a fixed separation $\vert{{\bf r}}_1-{{\bf r}}_2\vert$, with
\begin{equation}
    P({\bf r}) = P_1 \delta^2({\bf r}-{\bf r}_1) + P_2 \delta^2({\bf r}-{\bf r}_2),
    \label{gfdhjksa1}
\end{equation}
by simply superposing the solutions for the two single sources. This leads to the following contour plot of $u_z$ and $u_{x,y}$ in the $(z,x)$ (or $(z,y)$) plane, see Fig.~\eqref{two_points}. 

Solving the Navier equation Eq.~\eqref{navier} with a Green's function dyadic ${\cal G}_{ik}({\bf r} - {\bf r}')$, for details see Appendix~\eqref{appendix:0}, and taking into account that the external force acts only in the $z$ direction, one derives the elastic interactions free energy, in a completely analogous way as in the standard electrostatics, in the form
\begin{eqnarray}\nonumber
{\cal W}_{12}(\vert{{\bf r}}_1-{{\bf r}}_2\vert) &=& -{\textstyle\frac12} {\rho_s^2} \int_{S}\int_{S'} d^2S d^2S'~{\cal G}_{zz}(\vert{{\bf r}}-{{\bf r}}'\vert, z = z'= 0) P({{\bf r}}) P({{\bf r}}')\\
&=& -{\textstyle\frac12} {\rho_s^2} ~P_1 P_2~{\cal G}_{zz}(\vert{{\bf r}}_1-{{\bf r}}_2\vert, z_1=z_2=0),
\label{elastic_interact}
\end{eqnarray}
where $dS$ is the surface integral and ${\cal G}_{zz}(\vert{{\bf r}}-{{\bf r}}'\vert, z = z'= 0)$ is the $zz$ component of the dyadic Green function for $u_z$, so that we obtain from Eq.~\eqref{fund-navier} that the $z$-component of the total elastic deformation vector for the two point sources Eq.~\eqref{gfdhjksa1} is
\begin{eqnarray}
u_z ( r,z) = {\cal G}_{zz} (\vert {\bf r} - {\bf r}_{1}\vert, z) \rho_s P_{1} + {\cal G}_{zz} (\vert{\bf r} - {\bf r}_{2}\vert, z) \rho_s P_{2},  
\label{fund-navier1}
\end{eqnarray}
where the indices $1,2$ stand for either of the two point source in Eq.~\eqref{gfdhjksa1}.
${\cal G}_{zz}(\vert{{\bf r}}_1-{{\bf r}}_2\vert, z_1=z_2=0)$ entering the interaction free energy, Eq.~\eqref{elastic_interact}, can be obtained straightforwardly from the Love potential as
\begin{eqnarray}
{\cal G}_{zz}(r, z = z'= 0) &=& \nabla^{2} Z(r, z=0) - \frac{1}{2(1 - \sigma)}\frac{\partial^{2}Z(r,z=0)}{\partial z^{2}}.
\label{Greens}
\end{eqnarray}
where $Z(r,z=0)$ is the solution of Eq.~\eqref{eq:39} for a point-like surface force.

In the elastic interaction free energy, Eq.~\eqref{elastic_interact}, we excluded the self-interaction terms  that do not depend on the configuration of the sources. We refer to this interaction mediated by a finite slab of elastic material the {\sl Boussinesq interaction} between two point-like sources. In what follows we will generalize this to the Boussinesq interaction between two rigid rods. 

\begin{figure*}[t!]
\includegraphics[width=9cm]{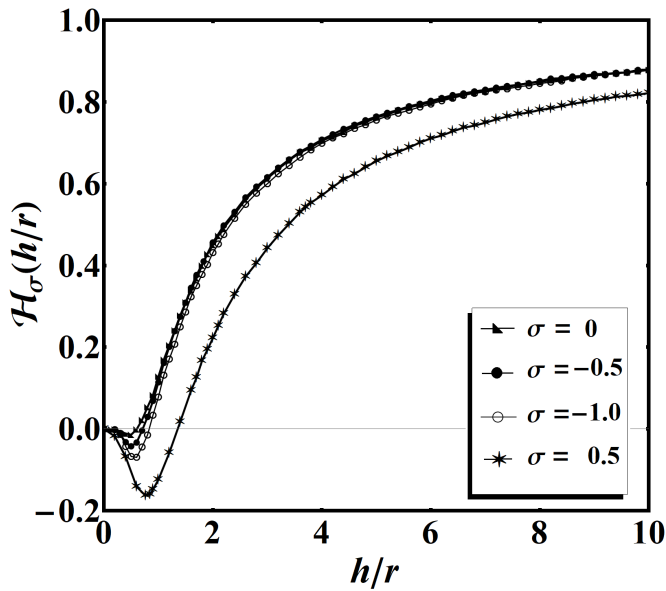}
\caption{The functional form of ${\cal H}_{\sigma}\left({h}/{r}\right)$ for four different values of $\sigma = 0, -0.5, -1.0, 0.5$ as marked in the legend. The functional dependence of ${\cal H}_{\sigma}({h}/{r})$ for small values of the argument changes sign at a small value of $h/r$, depending on $\sigma$, while for large values of the argument they all asymptote to $\lim_{x\longrightarrow\infty}{\cal H}_{\sigma}\left(x\right) = 1$. One observes that ${\cal H}_{\sigma}\left({h}/{r}\right)$ is effectively cut off for small enough values of the argument. This is the ``screening" effect of the finite thickness of the elastic matrix of the substrate.}
\label{FunH}
\end{figure*}

 The dependence of the Boussinesq interaction on the separation between the two point sources is straightforwardly analyzed in terms of the function  ${\cal H}_{\sigma}\left({h}/{r}\right)$, where $r = \vert{\bf r}_1 - {\bf r}_2\vert$, connected to ${\cal G}_{zz}(r,z=0)$, see Eq.~\eqref{hatzp1}, through
\begin{eqnarray}
{\cal G}_{zz}(r, z=0) &=& 
 \frac{2 (1-\sigma^2)}{2\pi E} \frac{1}{r} ~{\cal H}_{\sigma}\left(\frac{h}{r}\right).
\label{hatzp11}
\end{eqnarray}
The corresponding interaction free energy is then given by Eq. \ref{elastic_interact}, so that $\frac{1}{r} ~{\cal H}_{\sigma}\left(\frac{h}{r}\right)$ encodes all the spatial dependence of the interaction, while the prefactor ${ (\rho_s P_1 P_2)(1-\sigma^2)}/({\pi E})$, with the dimension of length, defines the strength of the interaction.
The dependence of  ${\cal H}_{\sigma}(x)$, where $x = h/r$, is shown in Fig.~\eqref{FunH}. Obviously ${\cal H}_{\sigma}(x)$ is a very slowly convergent function to its limit $\lim_{x \longrightarrow \infty} {\cal H}_{\sigma}(x) = 1$.
It is also evident that the vertical strain decreases with increase in the thickness of the substrate. Furthermore,  the vertical strain also decreases with the increase in the distance from the axis of the load in the lateral direction at any depth. From Eq.~\eqref{hatzp11} it follows that the Boussinesq interaction between two point surface sources is Coulombic for small separations but because of the finite thickness of the substrate it is cut off or screened for $r \geq h$. This long-range nature of the interaction makes it distinct from, {\sl e.g.},  the excluded volume interaction which is short range. In addition the interaction can change sign, implying attractive and repulsive regions as a function of the separation. Also, notice the difference between 2D electrostatic interaction that depends logarithmically on separation, and the Boussinesq interaction which is also effectively 2D but varies as the inverse for small separations.

\section{\label{sec:3}Boussinesq interaction between two rods}

\begin{figure}
\includegraphics[width=16cm]{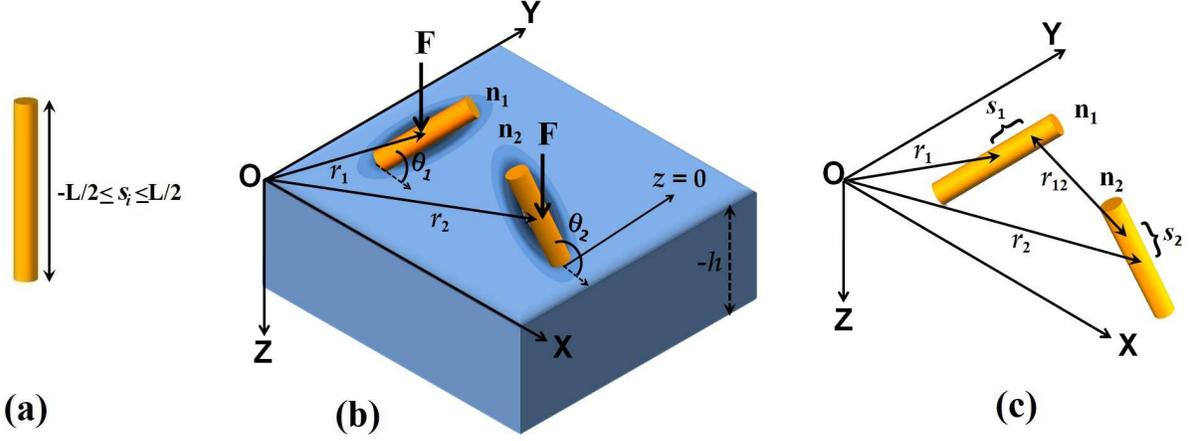}
\caption{Schematic presentation of the two-rod geometry. (a) $i$-th rod of length $L$ with the contour parameter $s_i$, (b) Coordinate geometry of the free surface of the substrate. (c). The two rods with their centers located at $ r_1 = (x_1, y_1)$ and ${r}_2 = (x_2, y_2)$ with respective orientations in the $(x,y)$ plane as ${\bf n}_1$ and ${\bf n}_2$.}
\label{two_rods}
\end{figure}

We now turn to rigid rods of length $L$ adsorbed onto the upper surface of the elastic substrate of finite thickness. An illustration of the instantaneous position and geometry of two rods is shown in Fig.~\eqref{two_rods}. The rods are identical in all aspects except for their orientation, and have by assumption a vanishing thickness. $p({\bf r})$ is now the linear density of the normal force acting along the rod and pointing downwards.  

Each rod, $i = 1,2$, is described by three parameters: the director, $\bf n$,  that defines the preferred orientation of the rod, assumed to be located in the plane perpendicular to the normal of the substrate surface, the center of mass located at ${\bf r}_i$, while $s_i$ is the contour parameter describing the position of the points along the rods in the frame of the rod itself, see the Fig.~\eqref{two_rods}, as
\begin{eqnarray}
{\bf R}_i({\bf n}_i, s_i) = {{\bf r}}_i + {\bf n}_i s_i, \qquad -{\textstyle\frac{L}2} \leq s_i \leq + {\textstyle\frac{L}{2}}.
\label{rod:position}
\end{eqnarray}
For $s_i = -{\textstyle\frac{L}2} \leq s_i \leq + {\textstyle\frac{L}{2}}$ the vector ${\bf R}_i({\bf n}_i, s_i)$ runs over all the points along the $i-th$ rod. The normal linear force density is then specified as 
\begin{eqnarray}
p(r) = p_1\delta^2({\bf r}-{\bf R}_1) + p_2\delta^2({\bf r}-{\bf R}_2).
\end{eqnarray}

Each two points along the two rods are now assumed to interact {\sl via} the Boussinesq interaction potential ${\cal W}_{12}(\vert{\bf R}_1({\bf n}_1, s_1)-{\bf R}_2({\bf n}_2, s_2)\vert)$. Following the analysis of Deutsch and Goldenfeld ~\cite{deutsch} for the case of charged rods with screened Coulomb interaction, as well as the analysis of Phan-Thien ~\cite{thein1979} for a slender rod-like body within an infinite elastic matrix, we use the principle of superposition for the Boussinesq interaction to calculate the total interaction free energy between the two rods by summing over the point-wise interactions along their lengths. This leads to the total interaction free energy as
\begin{equation}
{\cal W}_{12}({\bf n}_1, {\bf n}_2, \vert{\bf r}_1 - {\bf r}_2\vert) = - {\textstyle\frac12} \rho_s^2~p_1 p_2
 \int_{s_1}\int_{s_2} ds_1 ds_2 ~{\cal G}_{zz} (\vert{\bf R}_1({\bf n}_1, s_1) - {\bf R}_2({\bf n}_2, s_2)\vert; z_1 = z_2=0),
\label{elastic_interact1}
\end{equation}
where ${\cal G}_{zz}(r, z=0)$ is again the Green's function as introduced in the previous section, see Eq.~\eqref{Greens}. The above expression is a complicated integral over the contour parameters of the two rods. One could in principle carry on the analysis with this form of the interaction energy, but we choose a much more transparent development, based on the assumption that the separation between the rods is much larger then their length. This approximation is consistent with the fact that we will ignore the short-range orientational Onsager interaction and concentrate solely on the long-range interactions.

In this limit we can expand the bracket term of Eq.~\eqref{elastic_interact1} into a Taylor series, akin to a multipole expansion in electrostatics, up to the second order in the separation between the rods, yielding 
\begin{eqnarray}\nonumber
\int_{s_1}\int_{s_2} ds_1 ds_2 && ~{\cal G}_{zz} \Big(\vert{\bf R}_1({\bf n}_1, s_1) - {\bf R}_2({\bf n}_2, s_2)\vert; z_1 = z_2=0\Big) = L^2~{\cal G}_{zz}(\vert{\bf r}_1 - {\bf r}_2\vert ;z_1=z_2=0) \nonumber\\
&&+ \frac{1}{2!}\bigg[\frac{L^4}{12}(n_1 \nabla_{{\bf r}_1})(n_1 \nabla_{{\bf r}_1}) {\cal G}_{zz}(\vert{\bf r}_1 - {\bf r}_2\vert ;z_1=z_2=0) + \frac{L^4}{12}(n_2 \nabla_{{\bf r}_2})(n_2 \nabla_{{\bf r}_2}) {\cal G}_{zz}(\vert{\bf r}_1 - {\bf r}_2\vert ;z_1=z_2=0)\bigg] \nonumber \\ 
&& + \frac{1}{4!}\bigg[ \frac{L^6}{48}(n_1 \nabla_{{\bf r}_1})(n_1 \nabla_{{\bf r}_1})(n_2 \nabla_{{\bf r}_2})(n_2 \nabla_{{\bf r}_2}) {\cal G}_{zz}(\vert{\bf r}_1 - {\bf r}_2\vert ;z_1=z_2=0)] + \dots.
\label{integration}
\end{eqnarray}
as all the odd powers of $s_{1,2}$ in the Taylor expansion integrate out to zero. 
After evaluating all the derivatives of the Love potential explicitly, what we remain with is an expansion of the interaction free energy in the form
\begin{eqnarray}
{\cal W}_{12}({\bf n}_1, {\bf n}_2, \vert{\bf r}_1 - {\bf r}_2\vert) &=& V^{(0)}(\vert{\bf r}_1 - {\bf r}_2\vert) + V_{ij}^{(1)}(\vert{\bf r}_1 - {\bf r}_2\vert)n_{i1}n_{j1} + V_{ij}^{(1)}(\vert{\bf r}_1 - {\bf r}_2\vert)n_{i2}n_{j2} \nonumber\\ && + ~V_{ijkl}^{(2)}(\vert{\bf r}_1 - {\bf r}_2\vert)(n_{i1}n_{j1})(n_{k2}n_{l2}) + \dots
\label{bouss1}
\end{eqnarray}
where we introduced the following shorthand $r = \vert{\bf r}_1 - {\bf r}_2\vert$ and $\hat{\bf q}^{12} = ({\bf r}_1 - {\bf r}_2)/\vert{\bf r}_1 - {\bf r}_2\vert$. As the above interaction free energy is quadratic in the director $\bf n$, it describes the long-range {\sl nematic interaction}. Explicitly, by introducing ${\cal G}_{zz}(r) = {\cal G}_{zz}(\vert{\bf r}_1 - {\bf r}_2\vert ;z_1=z_2=0)$, we derive
\begin{subequations}
\begin{eqnarray}
V^{(0)}(r) =  -\frac{L^2}{2}\rho_s^2 ~p_1p_2 ~{\cal G}_{zz}(r)
\label{v0}
\end{eqnarray}
\begin{eqnarray}
V_{ij}^{(1)}(r) =  -\frac{L^4}{48}\rho_s^2 ~p_1p_2 ~\bigg(\frac{\partial^2 {\cal G}_{zz}(r)}{\partial r^2} - \frac{1}{r}\frac{\partial {\cal G}_{zz}(r)}{\partial r} \bigg)\hat{q}_i^{12}\hat{q}_j^{12} = V^{(1)}(r) ~\hat{q}_i^{12}\hat{q}_j^{12}
\label{v1}
\end{eqnarray}
\begin{eqnarray}\nonumber
 V_{ijkl}^{(2)}(r)& =&  -\frac{L^6}{2304}\rho_s^2 ~p_1p_2 ~\bigg(\frac{\partial^4 {\cal G}_{zz}(r)}{\partial r^4} -\frac{6}{r}\frac{\partial^3 {\cal G}_{zz}(r)}{\partial r^3} + \frac{15}{r^2}\frac{\partial^2 {\cal G}_{zz}(r)}{\partial r^2} -\frac{15}{r^3}\frac{\partial {\cal G}_{zz}(r)}{\partial r}\bigg)\hat{q}_i^{12}\hat{q}_j^{12}\hat{q}_k^{12}\hat{q}_l^{12}\\
 &=& V^{(2)}(r)~\hat{q}_i^{12}\hat{q}_j^{12}\hat{q}_k^{12}\hat{q}_l^{12}.
 \label{v2}
\end{eqnarray}
\end{subequations}

\begin{figure*}[t!]
\includegraphics[width=9cm]{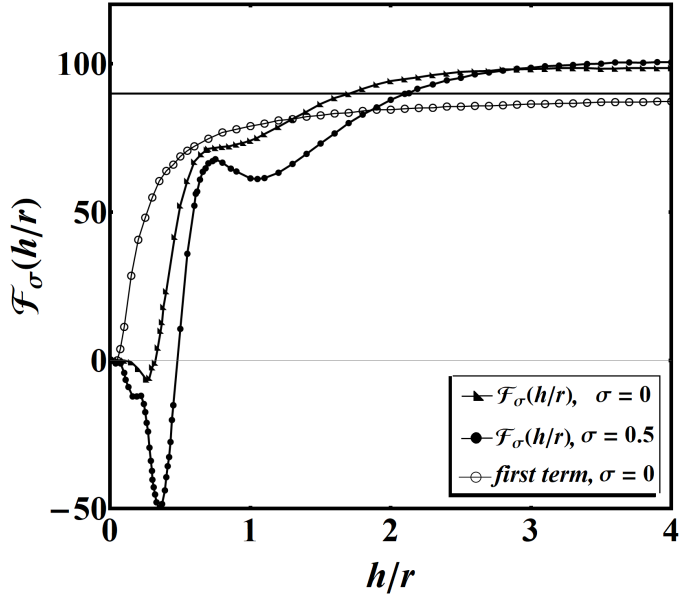}
\caption{The dimensionless scaling function ${\cal F}_{\sigma}(x)$ in the definition of the tensorial part of the rod-rod elastic interaction potential, Eq.~\eqref{hatzp20}, for two values of $\sigma$, and its first order term, see Eq. \eqref{hatzp20}. It exhibits monotonic and a superimposed  oscillatory components that can  change its sign for sufficiently small values of the argument (large value of the spacing $r$). Asymptotically $\lim_{x\rightarrow \infty}{\cal F}_{\sigma}(x) \longrightarrow 90$ for small value of the spacing $r$, indicated by the horizontal line. Clearly ${\cal F}_{\sigma}(x)$ exhibits a cutoff for small enough $x$.}
\label{Fsigma}
\end{figure*}

The interpretation of the three different terms in the expansion is straighforward: in Eq.~\eqref{v0} $V^{(0)}$ is the scalar interaction that depends only on the positions of the two rods but not on their orientations; in Eq.~\eqref{v1} $V_{ij}^{(1)}$ is the scalar-vector coupling interaction proportional to the square of the scalar product of the unit separation vector and the normal of the particle; in Eq.~\eqref{v2} $V_{ijkl}^{(2)}$ is the tensor interaction proportional to the product of the scalar products of the unit separation vector and the directors of the particles. The three terms together specify the elastic interaction free energy up to the order $(L/r)^4$ as the product $L p_i$ can be introduced as the total ``elastic charge" of the rod.

Since it will become clear that only the tensor interaction potential is relevant for the orientational ordering transition, taking into account Eq.~\eqref{hatzp11} we deduce from Eq.~\eqref{v2}
\begin{eqnarray}
V^{(2)}(r) &=& -\frac{2 (1-\sigma^2)}{2\pi E} \frac{\rho_s^2 ~p_1p_2~L^6}{2304} 
  \frac{1}{r^5} ~\left( 90 {\cal H}_{\sigma}(x) + 248 x{\cal H}_{\sigma}^{'}(x)+141 x^2{\cal H}_{\sigma}^{''}(x)+31x^3{\cal H}_{\sigma}(x)^{'''} + x^4{\cal H}_{\sigma}^{(iv)}(x)\right)  \nonumber\\
  &=& -\frac{2 (1-\sigma^2)}{2\pi E} \frac{\rho_s^2 ~p_1p_2~L^6}{2304} 
  \frac{1}{r^5} {\cal F}_{\sigma}(x), 
\label{hatzp20}
\end{eqnarray}
where $x = h/r$ and ${\cal H}_{\sigma}(x)$ has been defined in Eq.~\eqref{hatzp1}. From Fig.~\eqref{Fsigma} it is clear that ${\cal F}_{\sigma}(x)$ has a behavior that exhibits a monotonic and an oscillating component for a range of small values of $x$, the relative importance of the two dependent on $\sigma$. The oscillatory component can actually lead to negative values of ${\cal F}_{\sigma}(x)$. The orientational component of the interaction is thus non-monotonic for large lateral separation between the rods. In addition, for large values of the separation between the rods, $r \gg h$, the interaction is effectively cut off.

Analyzing the different components of ${\cal F}_{\sigma}(x)$ it is clear that apart from the first term they mostly cancel each other for large $x$, and that $\lim_{x\rightarrow\infty} {\cal F}_{\sigma}(x) = 90 \lim_{x\rightarrow\infty} {\cal H}_{\sigma}(x) = 90$. In any case, the strength of the tensorial Boussinesq interaction decays very steeply with the separation and is in addition cut off for large separations between the rods by the effects of the finite thickness of the substrate. 


\section{\label{sec:4}A system of $N$ identical rods}

\begin{figure}
\includegraphics[width=9cm]{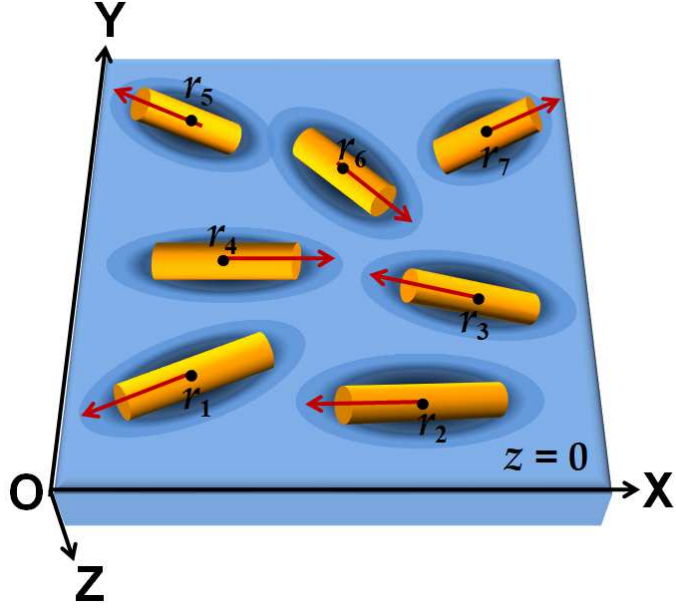}
\caption{A schematic diagram of rods on an elastic substrate. Each rod is associated with a deformation field around it, leading to mutual interactions between them. Rods with their centers located at ${r}_i = (x_i, y_i,)$ and with orientations in the $(x,y)$ plane (represented by red arrows).}
\label{nrod}
\end{figure}

In this section, we confine ourselves to a system of $N$ indistinguishable rigid rods, differing only with respect to their orientations and positions, deposited onto an elastic substrate of finite thickness. The rods are oriented randomly on the surface in the x-y plane at $z=0$, see Fig.~\eqref{nrod}. We assume that the interaction potential between rods is pairwise and is given by the Boussinesq interaction in the form valid for large separation between the rods, Eq.~\eqref{bouss1}.  The total elastic interaction energy between rods is then represented by
\begin{eqnarray}
{\cal W}(N) &=& \sum_{\mu \neq \nu = 1}^{N}V^{(0)}({\bf r}^\mu - {\bf r}^\nu) + 2 \sum_{\mu \neq \nu=1}^{N}V_{ij}^{(1)}({\bf r}^\mu - {\bf r}^\nu) \left(n_{i}^{\mu} n_{j}^{\mu}\right) 
+ \sum_{\mu \neq \nu=1}^{N}V_{ijkl}^{(2)}({\bf r}^\mu - {\bf r}^\nu)(n_{i}^{\mu} n_{j}^{\mu})(n_{k}^{\nu} n_{l}^{\nu}), 
\label{inten1}
\end{eqnarray}
where all the interactions depend solely on $V^{(i)}({\bf r}^\mu - {\bf r}^\nu) \equiv V^{(i)}(\vert{\bf r}^\mu - {\bf r}^\nu\vert)$. The Greek indices are used for particle identification and Latin indices are for Cartesian components of the vectors. The interaction potential is obviously complicated and we will introduce some additional approximations to simplify the analysis of the thermal properties of this system.

\subsection{Collective description}

Instead of writing the configurational energy of the system in terms of particle coordinates and orientations, we will introduce two collective order parameters: the scalar {\sl particle density order parameter}, $\rho({\bf r})$, and the traceless tensor {\sl nematic order parameter density}, $Q_{ij}({\bf r})$, where ${\bf r} = (x,y)$ is the position vector pertaining to the surface of the substrate. The two collective order parameters are defined as follows, 
\begin{eqnarray}
\rho({\bf r}) = \sum_{\mu=1}^{N}\delta^{2}({\bf r} - {\bf r}^\mu) \qquad {\rm and} \qquad Q_{ij}({\bf r}) = \sum_{\mu=1}^{N}\hat{Q}_{ij}^{\mu}\delta^2({\bf r}- {\bf r}^\mu), \qquad {\rm with} \qquad Q_{ii}({\bf r}) = 0
\label{ngjkrewl}
\end{eqnarray}
where the traceless quadrapolar nematic tensor order parameter for a two dimensional (2D) system is defined standardly as 
\begin{eqnarray}
\hat{Q}_{ij}^{\mu} = n_{i}^{\mu}n_{j}^{\mu} - {\textstyle\frac{1}{2}}\delta_{ij}.
\end{eqnarray}
One notes the missing factor $\textstyle\frac32$ in front of the first term in the above equation: this is due to the fact that the embedding space is 2D (surface of the substrate) and not 3D. 

With the two collective order parameters we can now write the pairwise interaction energy of the system, Eq. \ref{inten1},  in a succinct form 
\begin{eqnarray}\nonumber
{\cal W}(N) &=& {\cal W}(0) + {\textstyle\frac{1}{2}}\int d^2{\bf r} d^2{\bf r}'\rho({\bf r})U^{(0)}({\bf r} - {\bf r}')\rho({\bf r}') + \int d^2{\bf r} d^2{\bf r}'\rho({\bf r})U^{(1)}({\bf r} - {\bf r}')\hat{q}_{i}({\bf r}')\hat{q}_{j}({\bf r}')Q_{ij}({\bf r}') \nonumber\\
&+& {\textstyle\frac{1}{2}}\int d^2{\bf r} d^2{\bf r}'Q_{ij}({\bf r})\hat{q}_{i}({\bf r})\hat{q}_{j}({\bf r})U^{(2)}({\bf r} - {\bf r}')\hat{q}_{k}({\bf r}')\hat{q}_{l}({\bf r}')Q_{kl}({\bf r}').
\label{gfchjdwk}
\end{eqnarray}
Substituting the definitions of the collective order parameters, Eqs.~\eqref{ngjkrewl}, we are clearly back to the coordinate form Eq.~\eqref{inten1}.  
The first term in the above equation is the self-energy of the system defined as
\begin{eqnarray}
{\textstyle\frac{1}{2}}{\cal W}(0) = \sum_{\mu=1}^{N}V^{(0)}({\bf r}^\mu - {\bf r}^\mu) = N V^{(0)}(0),
\end{eqnarray}
which formally diverges, but we need to remember that we left out the short-range repulsions that would regularize it. The self-energy term corresponding to the Boussinesq interactions is completely analogous to the self-energy of the Coulomb system, and its only consequence is to renormalize the chemical potential~\cite{Coulomb}. The other terms in Eq.~\eqref{gfchjdwk} are obtained as 
\begin{eqnarray}
{\textstyle \frac{1}{2}}U^{(0)}({\bf r} - {\bf r}') = V^{(0)}({\bf r} - {\bf r}') + {\textstyle V^{(1)}}({\bf r} - {\bf r}') +  {\textstyle \frac{1}{4}}V^{(2)}({\bf r} - {\bf r}')
\end{eqnarray}
\begin{eqnarray}
{\textstyle\frac{1}{2}}U^{(1)}({\bf r} - {\bf r}') ={\textstyle 2V^{(1)}}({\bf r} - {\bf r}') + {\textstyle V^{(2)}}({\bf r} - {\bf r}')
\end{eqnarray}
\begin{eqnarray}
{\textstyle\frac{1}{2}}U^{(2)}({\bf r} - {\bf r}') = {\textstyle V^{(2)}}({\bf r} - {\bf r}')
\label{u2}
\end{eqnarray}
where $V^{(0)}$, $V^{(1)}$ and $V^{(2)}$ have been defined in Eqs.~\eqref{v0}, \eqref{v1} and \eqref{v2}, and depend only on the absolute value of the argument. In Eq.~\eqref{gfchjdwk} the various area integrals contain an orientational part and radial part:
\begin{eqnarray}
    \int d^2{\bf r} = 
    \int dS ~\Big< \dots \Big>_{\varphi} 
    \qquad {\rm where} \qquad \Big< \dots \Big>_{\varphi} \equiv \frac{1}{2\pi} \int_0^{\pi}\dots d\varphi,
    \label{bcfgsyku}
\end{eqnarray}
where $\varphi$ is the polar angle in cylindrical coordinates and $dS = 2\pi r dr$ is the area element. Clearly the polar angle integrals in Eq.~\eqref{gfchjdwk} can be evaluated explicitly as the orientational dependence of the interaction free energy is also explicit. This can be done by noticing that in 2D
\begin{eqnarray}
\langle\hat{q}_i\hat{q}_j\rangle _\varphi = {\textstyle\frac{1}{2}} \delta_{ij}
\end{eqnarray}
\begin{eqnarray}
\langle\hat{q}_i\hat{q}_j\hat{q}_k\hat{q}_l\rangle _\varphi  = {\textstyle\frac{1}{8}} (\delta_{ij}\delta_{kl} + \delta_{ik}\delta_{jl} + \delta_{il}\delta_{jk}).
\end{eqnarray}
Furthermore, we assume that the {\sl bulk state of the system is homogeneous}, so that the order parameter densities do not depend on the coordinate. Thus we finally remain with a much simplified form for the interaction energy
\begin{eqnarray}
\frac{{\cal W}(N) - {\cal W}(0)}{S} &=& {\textstyle\frac{1}{2}} a^{(0)}\rho^2 + {\textstyle\frac{1}{2}}  a^{(1)} \rho~ ({\rm Tr}~ Q_{ij}) + {\textstyle\frac{1}{16}} a^{(2)}  \Big(({\rm Tr}~ Q_{ij}))^2 + 2Q^2_{ij}\Big), 
\end{eqnarray}
where $S$ is the area of the undeformed elastic substrate and 
\begin{eqnarray}
\int dS ~ U^{(i)}({\bf r} - {\bf r}') = 2\pi a^{(i)} \qquad {\rm for} \qquad i=0,1,2,
\end{eqnarray}
where $a^{(i)}$ are the various orders of the second virial coefficient:  $a^{(0)}$ is the positional virial coefficient, $a^{(1)}$ is the positional-orientational coupling virial coefficient and $a^{(2)}$ is the orientational virial coefficient. It is the latter that will play a fundamental role for the orientational ordering transition in what follows.

\begin{figure*}[t!]
\includegraphics[width=9cm]{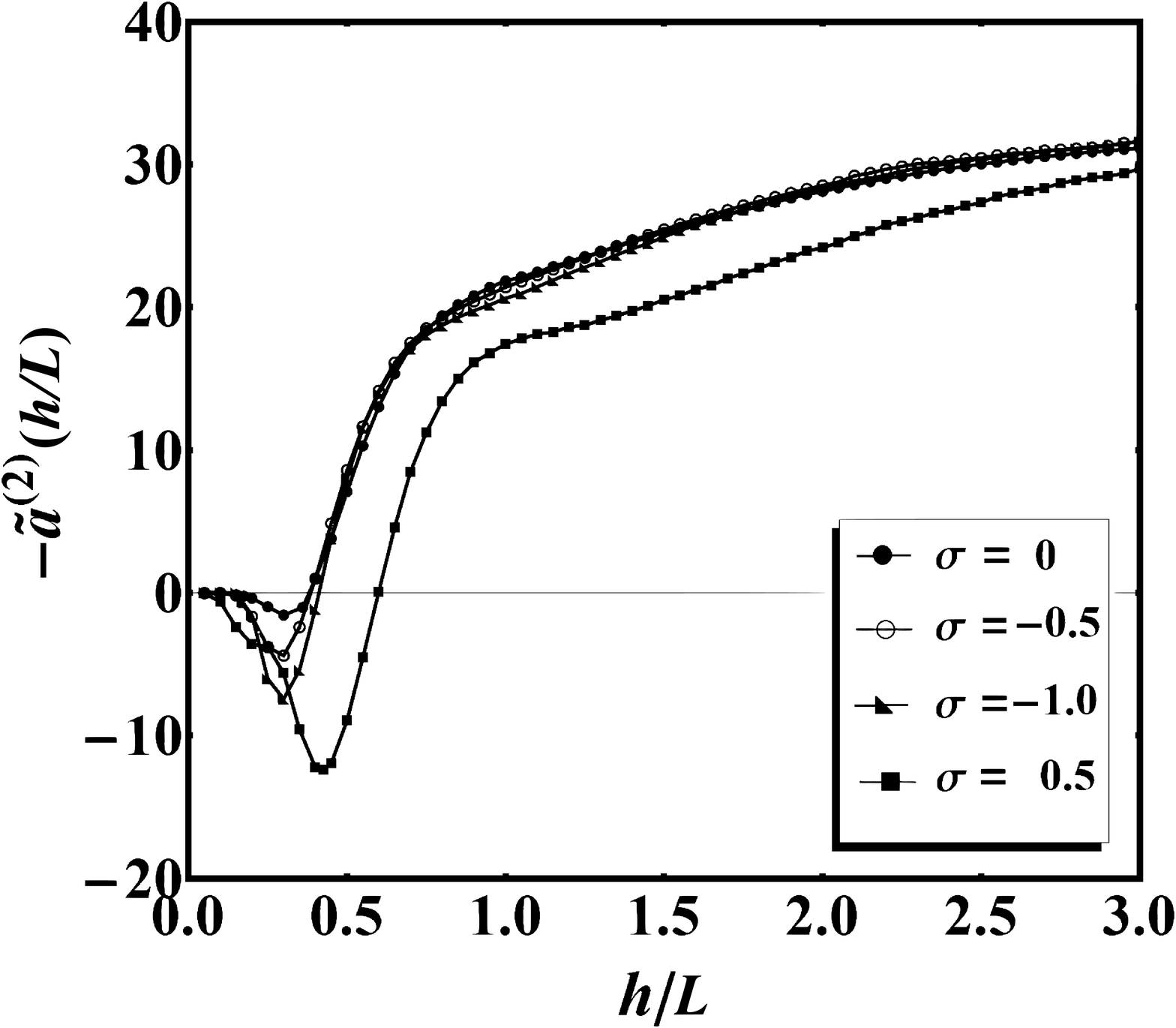}
\caption{The dependence of $\tilde{a}^{(2)}$ on the dimensionless variable $h/L$ for different values of $\sigma$. The non-monotonic dependence for small values of $h/L$, most pronounced for $\sigma = 0.5$ is not of physical relevance, as we are interested in the macroscopic limit $h/L \gg 1$. The limiting value $\lim_{x \longrightarrow \infty}\tilde{a}^{(2)}(x) = 30$ is independent of $\sigma$.}
\label{a2}
\end{figure*}

Taking into account that the nematic order parameter density is traceless, we finally remain with 
\begin{eqnarray}
\frac{{\cal W}(N) - {\cal W}(0)}{S}  = {\textstyle\frac{1}{2}} a^{(0)}\rho^2 + {\textstyle\frac{1}{8}} a^{(2)} Q_{ij}^2.
\label{f(n)}
\end{eqnarray}
This has the form consistent with the Lebwohl- Lasher interaction energy~\cite{BingSuiLu} and will much simplify later calculations. The orientational phase transition depends only on the second term  of the free energy and thus we only need to calculate $a^{(2)}$ which equals 
\begin{eqnarray}
2\pi a^{(2)} &=& \int dS ~ U^{(2)}({\bf r} - {\bf r}')  = {\textstyle\frac{16 \pi}{9}} \int_0^{\infty} r dr  ~V^{(2)}(r) = \nonumber\\
&&
= -\frac{2\pi }{2304}\rho_s^2 ~(p L)^2 {L^4}\int_0^{\infty} r dr  \bigg(\frac{\partial^4 {\cal G}_{zz}(r)}{\partial r^4} -\frac{6}{r}\frac{\partial^3 {\cal G}_{zz}(r)}{\partial r^3} + \frac{15}{r^2}\frac{\partial^2 {\cal G}_{zz}(r)}{\partial r^2} -\frac{15}{r^3}\frac{\partial {\cal G}_{zz}(r)}{\partial r}\bigg),
\label{bdfghjq}
\end{eqnarray}
where $r = \vert{\bf r} -{\bf r}'\vert$. We will show later that $a^{(0)}$ simply describes the van der Waals-like condensation transition at higher densities. In Appendix~\eqref{AppF} we derive the final form of the dimensionless second virial coefficient, based on Eq.~ \eqref{3bdfghjq13},

\begin{eqnarray}
\tilde{a}^{(2)} &=&  \left(\frac{L}{h}\right)^3 \int_0^{\frac{h}{L}} x^2 dx  ~{\cal F}_{\sigma}(x). 
\label{a2fin}
\end{eqnarray}
and ${\cal F}_{\sigma}(x)$, was defined in Eq.~\eqref{hatzp20}, while we assume that the shortest lengthscale in our model is proportional to the length of the rods, $\simeq L$. The orientational second virial coefficient, which will be the determining factor in our analysis of the orientational ordering transition below, then assumes the form, Eq.~\eqref{3bdfghjq13},
\begin{eqnarray}
a^{(2)} = \frac{8}{9} \frac{(1-\sigma^2)}{2304~\pi E} {\rho_s^2 ~(p L)^2~L} ~\tilde{a}^{(2)}
\end{eqnarray}

The non-monotonic structure of the second virial coefficient, see Fig.~\eqref{a2}, is a direct consequence of the effective orientational interaction ${\cal F}_{\sigma}(x)$, but is inconsquential in the macroscopic limit  of $h/L \gg 1$.

At this point it is illuminating to compare the present theory with the 2D problem of thin rods interacting only through short-range excluded volume interactions. The Onsager free energy can be conveniently expressed in the form Eq.~\eqref{f(n)} ~\cite{Doi1981}, where the 2D virial coefficient scales as $a^{(2)} \simeq k_BT~L^2$~\cite{Chen1993}.
Clearly the virial coefficient pertaining to the Boussinesq interactions is much larger for the same length of rods, and ignoring the Onsager interactions is thus entirely valid in this case. 

What we have accomplished in this section is to write the interaction energy due to Boussinesq elastic deformation of the substrate in the form containing only the collective order parameters. This leads naturally to the introduction of different second virial coefficients connected with the Boussinesq interaction, $a^{(0)}$ and $a^{(2)}$. From here we then need to proceed to the evaluation of the appropriate partition function of the canonical system.

\subsection{Canonical partition function and its field theoretical representation}

To determine the nature of the thermodynamic equilibrium state of a system consisting of $N$ rods, interacting {\sl via} the  Boussinesq elastic interactions, we must evaluate the corresponding canonical partition function defined as 
\begin{eqnarray}
{\cal Z}_N = \int \prod_{\mu =1}^{N} d^2{\bf r}_{\mu}~e^{-\beta{\cal W}(N)} = \int D[{\bf r}] e^{-\beta{\cal W}(N)} 
\end{eqnarray}
where $\beta = (k_BT)^{-1}$ is the inverse thermal energy and the product is over all $N$ rods and the integration is over all  configurations. We also dropped the indistinguishability term $N!$ as we will not be dealing with the condensation transition. We have shown that the interaction energy can be written in terms of collective order parameters ${\rho}({\bf r})$ and ${Q}_{ij}({\bf r})$, defined in Eq.~\eqref{ngjkrewl}, in the form
\begin{eqnarray}
{\cal W}(N) = {\cal W}\left[\rho({\bf r}), Q_{ij}({\bf r}) \right].
\label{fn}
\end{eqnarray}
We now introduce two decompositions of unity  
\begin{eqnarray}
1 = \int D[\rho({\bf r})] ~\delta(\rho({\bf r}) - \hat{\rho}({\bf r})) \qquad {\rm and} \qquad 1 = \int D[Q_{ij}({\bf r})]~ \delta(Q_{ij}({\bf r}) - \hat{Q}_{ij}({\bf r}))
\label{q1}
\end{eqnarray}
with the shorthand
\begin{eqnarray}
\hat{\rho}({\bf r}) = \sum_{\mu=1}^{N}\delta^2({\bf r} - {\bf r}_{\mu}); ~~~ \hat{Q}_{ij}({\bf r}) = \sum_{\mu=1}^{N}\left( n_i({\bf r})n_j({\bf r}) - \frac{1}{2}\delta_{ij}\right) \delta^2 ({\bf r} - {\bf r}_{\mu}),
\end{eqnarray}
and cast the functional delta function with its integral representation
\begin{eqnarray}
\delta\left(\rho({\bf r}) - \hat{\rho}({\bf r})\right) &=& \int D[\phi({\bf r})]~e^{-i \int d^2{\bf r}~\phi({\bf r})(\rho({\bf r}) - \hat{\rho}({\bf r}))} \nonumber\\
\delta\left(Q_{ij}({\bf r}) - \hat{Q}_{ij}({\bf r})\right) &=& \int D[\phi_{ij}({\bf r})]~e^{-i\int d^2{\bf r}~\phi_{ij}({\bf r})(Q_{ij}({\bf r}) - \hat{Q}_{ij}({\bf r}))},
\label{dq}
\end{eqnarray}
where we designate $\phi({\bf r})$ as the {\sl scalar molecular field} (note that in the case of Coulomb fluids this would be the fluctuating electrostatic potential, see Ref.~\cite{Coulomb}) and $\phi_{ij}({\bf r})$ as the nematic {\sl tensor molecular field}. We can then rewrite the partition function as 
\begin{eqnarray}
{\cal Z}_N &=& \int D[{\bf r}] e^{-\beta{\cal W} [\rho({\bf r}), Q_{ij}({\bf r})]} \times \int D[\rho({\bf r})] D[\phi({\bf r})] e^{-i\int d^2{\bf r}~\phi({\bf r})(\rho({\bf r})-\hat{\rho}({\bf r}))} \nonumber\\
&& ~~~~~~~~~~~~~~~~ \times \int D[Q_{ij}({\bf r})] D[\phi_{ij}({\bf r})] e^{-i\int d^2{\bf r}~\phi_{ij}({\bf r})(Q_{ij}({\bf r})-\hat{Q}_{ij}({\bf r}))}.
\end{eqnarray}

This formidable looking expression can be reworked a bit to yield
\begin{eqnarray}
{\cal Z}_N = \int D[\rho({\bf r})] D[\phi({\bf r})] D[Q_{ij}({\bf r})] D[\phi_{ij}({\bf r})]~e^{-\beta{\cal W}[\rho({\bf r};Q_{ij}({\bf r})]-i\int d^2{\bf r}~\phi({\bf r})\rho({\bf r}) -i\int d^2{\bf r}~\phi_{ij}({\bf r})Q_{ij}({\bf r})} \times \tilde{{\cal Z}}_N
\end{eqnarray}
where we introduced the partition function of the system in an scalar and tensor external field as
\begin{eqnarray}
\tilde{{\cal Z}}_N = \int \prod_{\mu=1}^{N}d^2{\bf r}_{\mu}~ e^{i\sum_{\mu=1}^{N} \phi({\bf r}_\mu) +  i\sum_{\mu=1}^{N}(\phi_{ij}({\bf r}_\mu) ~n_in_j-\frac{1}{2}Tr\phi_{ij}({\bf r}))} = \left[ \tilde{{\cal Z}}_1\right]^N
\end{eqnarray}
as the trace is diagonal and the single particle partition function in the external fields is then 
\begin{eqnarray}
\tilde{{\cal Z}}_1  = \int d^2{\bf r} ~ e^{i(\phi({\bf r})-\frac{1}{2}Tr \phi_{ij}({\bf r}) + \phi_{ij}({\bf r}) n_in_j)}. 
\label{tildeq}
\end{eqnarray}
From here
\begin{eqnarray}
\tilde{{\cal Z}}_N = \int D[\rho({\bf r})] D[\phi({\bf r})] D[Q_{ij}({\bf r})] D[\phi_{ij}({\bf r})]~ e^{-H[\rho({\bf r})\phi({\bf r})Q_{ij}({\bf r})\phi_{ij}({\bf r})]}
\label{bdgjekqw}
\end{eqnarray}
with the effective field action defined as
\begin{eqnarray}
H[\rho({\bf r}),\phi({\bf r}), Q_{ij}({\bf r}),\phi_{ij}({\bf r})] =  \beta{\cal W}[\rho({\bf r}), Q_{ij}({\bf r})] + i\int d^2{\bf r} ~\phi({\bf r}) \rho({\bf r}) + i\int d^2{\bf r}~ \phi_{ij}({\bf r}) Q_{ij}({\bf r}) - N \ln{\tilde{{\cal Z}}_1}.
\label{hpq}
\end{eqnarray}
This is the final form of the canonical partition function for a system of $N$ rods interacting via the Boussinesq interaction. We note here that the derivation of the field representation of the partition function follows closely the related developments in the field representation of the Coulomb fluid partition function \cite{Coulomb}, the main difference being that the orientational part of the Boussinesq interactions between rods brings fourth not only the scalar collective order parameter but also the tensorial collective order parameter.

Since the field action is highly non-linear there is no direct way to evaluate Eq.~\eqref{hpq}. Further approximations are therefore necessary. Just as in the case of the Coulomb fluid, where the saddle point approximation leads to the well known Poisson-Boltzmann theory, the saddle point approximation of 
Eq.~\eqref{bdgjekqw} will naturally yield an equivalent of the Maier-Saupe theory. In the next section we will see how.

\subsection{The saddle-point approximation and the Maier-Saupe theory}

We first invoke the approximation of a homogeneous ``bulk", meaning that the collective order parameters do not depend on the position along the surface. By invoking Eq.~\eqref{f(n)} and Eq.~\eqref{hpq} this yields the field action as

\begin{eqnarray}
H[\rho,\phi, Q_{ij},\phi_{ij}] = S\bigg(\frac{1}{2}a^{(0)}\rho^2 + \frac{1}{8}a^{(2)}Q_{ij}^2 + i~\phi\rho + i~\phi_{ij}Q_{ij} - \frac{N}{S} ln \tilde{{\cal Z}}_1\bigg),
\end{eqnarray}
where, to cut down on the clutter, we expressed the virial coefficients in terms of the thermal energy. In addition we can write (see Eq.~\eqref{tildeq}) 
\begin{eqnarray}
\tilde{{\cal Z}}_1 = 
S~e^{i\phi}\Big< e^{i\phi_{ij}(n_in_j - \frac{1}{2}\delta_{ij})}\Big>_\varphi,
\end{eqnarray}
where the orientational average has been defined before, see Eq.~\eqref{bcfgsyku}. The saddle-point approximation is now defined as the stationary point of the field action, leading to the following two scalar equations

\begin{subequations}
\begin{eqnarray}
\frac{\delta H}{\delta \rho} = 0 \quad \longrightarrow \quad a^{(0)} \rho + i\phi = 0 \qquad {\rm and} \qquad \frac{\delta H}{\delta \phi} = 0 \quad \longrightarrow \quad i\rho - \frac{N}{S}\frac{\partial}{\partial \phi}\log\tilde{{\cal Z}}_1 = 0.
\label{four1}
\end{eqnarray}
Elaborating these two equations further leads to the van der Waals theory of condensation~\cite{widom82}, which usually takes place at higher concentrations then the orientational phase transition. We will not delve into the details of the former as it is less interesting than the latter.
\text{The corresponding two tensorial saddle-point equations are obtained as}
\begin{eqnarray}
\frac{\delta H}{\delta Q_{ij}} = 0 \quad \longrightarrow \quad \frac{1}{4}a^{(2)}Q_{ij} + i\phi_{ij} = 0 \qquad {\rm and} \qquad \frac{\delta H}{\delta \phi_{ij}} = 0 \quad \longrightarrow \quad iQ_{ij} - \frac{N}{S}\frac{\partial}{\partial \phi_{ij}}\log{\tilde{{\cal Z}}_1} = 0.
\label{four2}
\end{eqnarray}
\end{subequations}
From the first pair of Eq.~\eqref{four1}, we see that the molecular field $\phi$ at the saddle point is purely imaginary, $\phi \longrightarrow i \phi$, just as in the case of the saddle-point approximation for Coulomb fluids~\cite{Coulomb}. The second identity then implies
\begin{eqnarray}
i\rho - i\frac{N}{S}\tilde{\cal Z}_{1}^{-1}\frac{\partial \tilde{{\cal Z}}_1}{\partial \phi} = i\rho - i \frac{N}{S} = 0 \quad \longrightarrow \quad \rho = \frac{N}{S}= c_R,
\end{eqnarray}
where $c_R$ is the surface density of the rods (notably different from $\rho_s$, the density of the elastic matrix). Let's concentrate on Eq.~\eqref{four2} then. The first equation implies that at the saddle point $\phi_{ij} \longrightarrow i \phi_{ij}$, so that it is also purely imaginary.  The second equation, after cancelling the imaginary unit, then becomes
\begin{eqnarray}
Q_{ij} = \frac{N}{S}\frac{\Big< (n_{i}n_{j} - \frac{1}{2}\delta_{ij})e^{-\phi_{ij}(n_{i}n_{j} - \frac{1}{2}\delta_{ij})}\Big>_\varphi}{\Big< e^{- \phi_{ij}(n_{i}n_{j} - \frac{1}{2}\delta_{ij})}\Big>_\varphi} = c_R~ \Big<\!\!\!\Big< n_in_j -\frac{1}{2}\delta_{ij}\Big>\!\!\!\Big>_{\varphi},
\label{qnv}
\end{eqnarray}
where we introduced the orientational average
\begin{eqnarray}
\Big<\!\!\!\Big< n_in_j -\frac{1}{2}\delta_{ij}\Big>\!\!\!\Big>_{\varphi} = \frac{\int_0^{\pi} d\varphi~(n_in_j - \frac{1}{2}\delta_{ij}) e^{-\phi_{ij}(n_in_j - \frac{1}{2}\delta_{ij})}}{\int_0^{\pi} d\varphi~ e^{-\phi_{ij}(n_in_j - \frac{1}{2}\delta_{ij})}}.
\end{eqnarray}

\begin{figure}
\includegraphics[width=17cm]{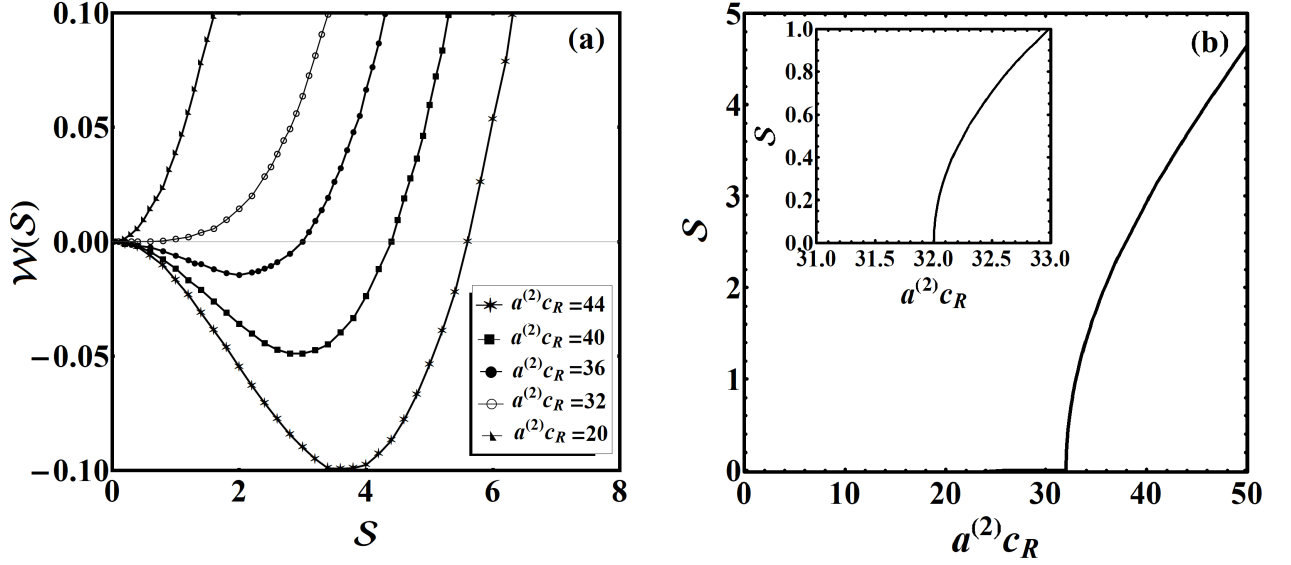}
\caption{(a). Free energy Eq.~\eqref{freen} as a function of the order parameter ${\cal S}$ for different values of $a^{(2)} c_R = 44, 40, 36, 32, 20$. (b). Order parameter ${\cal S}$ as a function of $a^{(2)}c_R$ from the Maier-Saupe equation Eq.~\eqref{dawson}.  At $(a^{(2)}c_R)^*=32.04$ there is a continuous order-disorder transition. The inset shows details around the critical point.}
\label{MSeq}
\end{figure}
In Eq.~\eqref{qnv} we recognize the Maier-Saupe approximation that describes the equilibrium properties of the system. We now assume that the saddle-point solution of the molecular field $\phi_{ij}$, from the first identity of Eq.~\eqref{four2}, has the form
\begin{eqnarray}
\phi_{ij} = \frac{1}{4}a^{(2)}Q_{ij} = {\cal S}~ \Big(\hat{n}_i \hat{n}_j - \frac{1}{2}\delta_{ij}\Big)
\end{eqnarray}
where $\cal S$ is {\sl proportional} to the orientational order parameter, the proportionality coefficient being $\frac{1}{4}a^{(2)}$, which quantifies how well-aligned the molecules are, and $\hat{\bf n}$ is the director of the ordered phase. Therefore
\begin{eqnarray}
Q_{ij} = \frac4{a^{(2)}} \phi_{ij} = c_R~ \Big<\!\!\!\Big< n_{i}n_{j}-\frac{1}{2}\delta_{ij} \Big>\!\!\!\Big>_{\varphi}.
\label{phai}
\end{eqnarray}
Multiplying this with $(\hat{n}_i \hat{n}_j - \frac{1}{2}\delta_{ij})$ on both sides, taking into account that $\hat{\bf n}\cdot{\bf n} = \cos{\varphi}$ and contracting over the tensorial indices leads us to
\begin{eqnarray}
 {\cal S} + {\textstyle\frac{1}{4}}~\vert a^{(2)}\vert  c_R~\frac{\int_0^{\pi} (\cos^2 \varphi - \frac{1}{2})e^{-S(\cos^2 \varphi - \frac{1}{2})}d \varphi}{\int_0^{\pi}e^{-S(cos^2\varphi - \frac{1}{2})}d\varphi} = 0,
\label{dawson}
\end{eqnarray}
where we took into account that the orientational second virial coefficient is negative. Except for the dimensionality, 2D vs. 3D, this equation is equivalent to the Kleinert formulation of the Maier-Saupe theory~\cite{Kleinert}. As it describes a lyotropic transition, the strength of nematic interaction quantified by the Boussinesq second  virial coefficient is multiplied by the density of the rods, $c_R$, {\sl i.e.} the larger the local density of the rods, the smaller needs to be the orientational interaction driving the isotropic-nematic transition. Notably, the above solution is just a minimum of an effective free energy
\begin{eqnarray}
{\cal W}({\cal S}) = - \frac{2}{\vert a^{(2)}\vert c_R} ~{\cal S}^2 + \log{\int_{0}^{\pi}e^{-S(cos^2 \varphi - \frac{1}{2})}d\varphi} =  -\frac{2}{\vert a^{(2)}\vert c_R} ~{\cal S}^2 + \log{I_0({\cal S}/2)},
\label{freen}
\end{eqnarray}
where we took into account the integral representation of the modified Bessel function of the first kind $I_0(z)$. The solution of equation Eq.~\eqref{dawson}, see Fig.~\eqref{MSeq}, exhibits a continuous ordering transition at a critical value of $(a^{(2)}\rho)^* \simeq 32.04$, characterized by a non-vanishing value of $\cal S$. Interestingly enough, the only formal difference between the 2D and the 3D cases is that instead of $I_0({\cal S}/2)$ in Eq.~\eqref{freen}, one ends up with ${\rm Exp}({-{\textstyle\frac13} {\cal S})} {\cal D}(\sqrt{{\cal S}}/\sqrt{{\cal S}}$, where ${\cal D}(x)$ is the Dawson function~\cite{Kleinert}. However the solution of the Maier-Saupe equation then leads to a first order transition, with a finite jump in ${\cal S}$.


\section{\label{sec:5}Discussion}
 In this work we studied theoretically the orientational ordering of anisotropic rigid rods deposited onto the surface of a soft elastic substrate of finite thickness. The substrate is assumed to be a homogeneous and isotropic elastic medium. The spatial profile of a deformed substrate and the corresponding elastic displacements due to adsorbed rods are calculated within the linear {\sl Boussinesq theory} for a finite thickness substrate. In this model the rods are represented by a concentrated force normal to the substrate surface, uniformly distributed over their length with a constant linear density. The long-range, Coulomb-like effective interactions between two rods, mediated by the elastic displacement of the substrate, are then pairwise additive on the linear level. At the macroscopic length scale we formulate the partition function of a canonical ensemble of rods interacting {\sl via} these substrate elastic deformation mediated Boussinesq interactions and derive a mean-field type theory, yielding a  continuous isotropic-nematic transition. 

 The model of a constant normal linear force density along the surface-deposited thin rods, similar to the model of rod-like polyelectrolytes with uniform linear charge density~\cite{deutsch}, is a coarse grained and simplified version of the physical reality in the sense that, microscopically, there is an adhesion energy distributed over the area of the finite diameter rod in contact with the substrate. The local substrate surface deformation is then very similar to the case of adhesion of colloids to a soft membrane~\cite{Deserno_2002, weikl2003,Deserno_2003, Weikl__2012} or wrapping of colloids along an elastic filament~\cite{dean2012} and could be computed in the same way. In scaling terms the product of the normal force  and the surface depression of the substrate surface  should be proportional to the product of the adhesion (wrapping) surface energy density times the contact surface area. While in principle the calculation of effective rod-rod interactions with explicit adhesion to the surface is possible, it would make the analysis of a canonical ensemble of such rods unfeasible.

We first restrict ourselves to interactions between two point-like surface forces acting at ${\bf r}_1$ and ${\bf r}_2$, respectively, and discuss the influence of substrate thickness on the corresponding strain fields in Sec.~\eqref{sec:2}. The dependence of the vertical strain $u_z({\bf r},z)$ on the substrate thickness is plotted in Fig.~\eqref{uzprofile}, where different curves correspond to different values of the thickness, $h( = 12,6,3)$. We set the Poisson ratio $\sigma = 0.5$ in all three cases. The surface pressure on the top surface of the substrate, $z = 0$, is set by the normal force density distribution of two point like surface forces, while deformation vectors at the bottom surface are constrained to  vanish at $z + h = 0$. We found that the vertical strain decreases with increasing substrate thickness, while for a fixed thickness, it decreases with the radial distance, $r$. Importantly, it can be seen that as the substrate thickness decreases, the iso-deformation contours shrink towards the $z =0 $ surface. For example, for the coordinate $z$ = -1.5 and $r$ = 0 the $u_z(r, z)$ increases by a factor of 0.85$P/(2\pi E)$ for $h$ = 3 and 1.68$P/(2\pi E)$ for $h$ = 12, which implies that the iso-deformation contour contraction for $h = 3$ is significantly larger (about 50$\%$) than that for $h=12$. Fig.~\eqref{two_points} depicts the iso-deformation contours of $u_z(r, z)$ and $u_{x,y}(r, z)$ due to the two point-like sources at dimensionless separation $\vert {\bf r}_1 - {\bf r}_2\vert =4$ for $\sigma$ = 0.01. Because on the Boussinesq level the problem is linear, the deformation fields around each point-like source force are obtained from a superposition principle and in turn result in an effective elastic attraction between the sources. The situation is similar to electrostatics, the differences stemming from the fact that electrostatics is a scalar theory while elasticity is a tensor (vector) theory, thus the change in sign in the interaction. 

It is obvious from Fig.~\eqref{two_points} that the thinner the substrate, the higher the value of the  strain. The separation dependence of the Boussinesq interaction between two point sources as a function of the separation between them is investigated in Sec.~\eqref{sec:2}. Fig.~\eqref{FunH} shows the dependence of the interaction scaling function  ${\cal H}_{\sigma}\left({h}/{r}\right)$ on $h/r$ for four different values of Poisson ratio $\sigma$ =(0.5, 0.01, -0.5, -1.0). Clearly, ${\cal H}_{\sigma}\left({h}/{r}\right)$ saturates asymptotically to 1 as $h/r \rightarrow \infty$ and approaches 0 for $r >> h$. This yields an effective Boussinesq interaction decaying as the inverse power of the separation just as in 3D electrostatics. For smaller value of $h/r$, dramatic changes are observed, with ${\cal H}_{\sigma} \left({h}/{r} \right)$ exhibiting a cut-off dependent on the thickness of the substrate as well as a change of sign of the interaction. The cut-off seems to describe a ``screening" effect of the finite thickness of the substrate, meaning that if the two point surface sources are close enough the interaction between them is Coulombic but is then effectively cut off for $r \geqslant h$. 

In Sec.~\eqref{sec:3}, we discuss the substrate deformation caused by two adsorbed rods as well as the effective Boussinesq elastic interaction between them, as a function of their separation and orientation. Fig.~\eqref{Fsigma}, shows the variation of the orientational nematic interaction in terms of the scaling function ${\cal F}_{\sigma}\left({h}/{r}\right)$. We notice that for small separation between rods ($r << h$), ${\cal F}_{\sigma}\left({h}/{r}\right)$ asymptotically approaches a constant value. In general, ${\cal F}_{\sigma} \left( {h}/{r} \right)$ is composed of a monotonic and a superimposed  oscillatory component that yields a negative interaction energy if the rods are away from each other (large value of $r$). It is clear that the rod-rod interaction decays sharply with the separation between the rods, and in fact for small enough $h/r$, ${\cal F}_{\sigma}\left({h}/{r}\right)$ limits to 0. It is worth mentioning that the material properties of the substrate have also a significant impact on the interaction between surface inclusions mostly through the Poisson ratio. As shown in Fig.~\eqref{Fsigma}, the non-monotonic interaction component of ${\cal F}_{\sigma}\left( {h}/{r} \right)$ is more pronounced for $\sigma = 0.5$ as compared to $\sigma = 0$.

In Sec.~\eqref{sec:4}, we extend our model to a canonical thermodynamic ensemble of $N$ rods. The interaction potential between rods is assumed to be pairwise and is given by the Boussinesq interaction, leading to elastic interaction energy between rods given by Eq.~\eqref{inten1}. We investigate the critical behavior of the orientational order parameter and the free energy through a solution of the Maier-Saupe equation that we derive analytically and exactly. For different values of the product of the orientational virial coefficient and the macroscopic surface density of the rods, $a^{(2)}c_R$ in Eq.~\eqref{freen}, as illustrated in Fig.~(\ref{MSeq}(a)), we conclude that for higher values of $a^{(2)}c_R$, there is a continuous lyotropic transition from an isotropic to a nematic phase. This happens at the critical value $a^{(2)} c_R = 32.04$, at which the orientational order parameter decreases sharply but continuously to zero, implying a second order transition. In order to have a nematic surface phase, according to Eq. \ref{3bdfghjq13}, the elastic parameters of the system need to satisfy 

\begin{eqnarray}
\frac{5}{432 \pi ~} \frac{(1-\sigma^2)}{E ~k_BT} F_s^2~L ~c_R \geq 32.04.
\label{vafejtys}
\end{eqnarray}
where $F_s = \rho_s P = \rho_s (pL)$ is the total normal force per rod that can be interpreted as an ``elastic charge" of the rod. While the value of this force is a phenomenological parameter of our model, the above limit is fulfilled even if it is very small. In fact taking reasonable values for the Young modulus of a gel, length of the rod on the order of 100 nm, 
the inequality Eq. \ref{vafejtys}
can be easily fulfilled even for small surface concentrations of rods. We can also investigate the ratio between the Onsager excluded volume interaction in 2D and the orientational virial coefficient \cite{Kayser1978,Chen1993,Frenkel2000}. This ratio scales as

\begin{eqnarray}
 \frac{(1-\sigma^2)}{E~k_BT} \frac{F_s^2}{L},
\end{eqnarray}
being much larger then 1 for reasonable values of the parameters. This also implies that the long-range Boussinesq elastic interactions are more important for surface ordering on soft substrates then the standard short-range (Onsager) steric effect. 

At higher values of $a^{(2)}c_R$ (dense system) the stable phase is thus a nematic phase and the transition to this nematic phase is lyotropic at constant temperature. This second order lyotropic transition is purely an effect of a 2D nature of the problem, or a quasi 2D nature, as the interactions potential is fully 3D. In fact, the same analysis for a complete 3D orientational ordering would yield a first order transition, see \cite{Kleinert}.

These results are in line with previous studies of continuous isotropic-nematic transition in 2D ~\cite{Frenkel2000, denham1980, frenkel, Lago2003, heras13, lomba2005, geng2009, small17, zonta2019}. A continuous transition was predicted by the 2D version of the Maier-Saupe theory for cylinders interacting via a weak anisotropic pair potential~\cite{denham1980} and for particles in strong confinement~\cite{lomba2005,heras13}. Even small changes in density and shape of anisotropic particles dramatically tune both structure and transition properties of mesophases~\cite{Frenkel2000,frenkel,Lago2003,geng2009,zonta2019}.
As example, a new type of 2D nematic, a tetratic  phase was hypothesized, by using an interaction of the Maier-Saupe type with terms representing both twofold and fourfold anisotropic interactions~\cite{geng2009}. It was found that the isotropic-tetratic transition is always second-order and independent of the symmetry breaking parameter $\kappa$ (the ratio between fourfold and twofold interactions). The tetratic-nematic transition is second-order for small $\kappa$, but becomes first-order at the tricritical point~\cite{geng2009}. In a recent work, a first-order isotropic-nematic transition is observed for intermediate aspect ratio and a small anisotropic interaction parameter, while a continuous isotropic-tetratic transition is found for small aspect ratio and small anisotropic interaction parameter, followed by a continuous tetratic-nematic transition at higher densities~\cite{zonta2019}. However, in simulations of hard rods, no tetratic phase was found~\cite{Frenkel2000,Lago2003}. They found, isotropic, nematic, smectic and full crystalline phase that depend upon concentration and aspect ratio~\cite{Frenkel2000,Lago2003}. 
The importance of elongated particles in the context of Isotropic-nematic transition in 2D has been demonstrated for a wide range of systems, including assembly of rod-like virus on polymer surface~\cite{yoo2006}, and self-organization of amyloid fibrils at fluid interfaces~\cite{jordens2013}. Moreover, particle shape can induce local interfacial deformations with strong anisotropic capillary forces between particles at liquid-air interface~\cite{loudet2005}. In this system, long-range interactions are governed by surface tension, while in our model the full elastic deformation of the substrate comes into play.

We note that the saddle-point Maier-Saupe theory is of course approximate and other approaches could be used to go beyond its limitations, possibly also yielding a more accurate description of this lyotropic surface ordering. There is, however, no systematic evaluation of different approaches to the charged rod ordering transition and we can only list a few of the possible alternatives. The already cited work of Deutsch and Goldenfeld\cite{deutsch, Deutsch82} for thin charged rods already presents an alternative collective coordinate transformation method applied to the ordering of charged rods in 3D. In addition, for thin rod-like charged cylinders a generalized Onsager theory could also be used to calculate the ordering transition~\cite{Odijk86, Trizac14, Jho14}, while a generalized variational field theory~\cite{Lue06} and an order parameter based mean-field approximation~\cite{Muthu99} both lead to a first order transition in 3D.

In conclusion, we would like to emphasise that the deformations and the mechanical response of elastic substrates significantly alters the directionality of the adsorbed rod-rod interactions and strongly influences their phase behavior. Throughout this work, we aimed to understand the governing parameters that tune the phase behavior of adsorbed rods on soft elastic surfaces. These results are therefore likely to have a significant impact on the design of smart surfaces or diagnostic tools for the detection of pathogenic molecules such as those based on surface binding of rod-shaped viruses and bacteria. Our study can be used as a model system to understand their biological counterparts. For example, responses of soft tissue to surface elastic forces, or surfaces of hydrogels are just a few to name. In these cases a dynamic generalization of our model, possibly along the lines of the recent work by Bar-Haim and Diamant \cite{Diamant__2020}, would yield a rich enough model to be able to describe that kind of phenomenology. Our results may even help to extend previous studies on living cells on an elastic substrate where the forces can be induced actively by the living cells themselves, typically modeled by anisotropic force contraction dipoles ~\cite{schwarz2002,bischofs2004,yuval2013}.

\section{Data Availability Statement}

The data that support the findings of this study are available from the corresponding author upon reasonable request.

\section{Acknowledgement}

SK and RP would like to acknowledge the support of the 1000-Talents Program of the Chinese Foreign Experts Bureau, of the School of physics, University of the Chinese Academy of Sciences, Beijing and of the Institute of physics, Chinese Academy of Sciences, Beijing. FY acknowledges the support of the National Natural Science Foundation of China (Grant No. 11774394) and the Key Research Program of Frontier Sciences of Chinese Academy of Sciences (Grant No. QYZDB-SSW-SYS003). We would like to express our gratitude to Prof. H. Diamant for his comments on an earlier version of this MS.


\nocite{*}
\bibliography{Manuscript.bib}


\appendix
\section{\label{appendix:0} Navier equation and its Green's dyadic}

We start with minimization of the elastic free energy in the presence of external forces, see Ref.~\cite{LL} for details, that implies the Navier equation, Eq.~\eqref{navier}, as
\begin{eqnarray}
\nabla ^{2}\vct{u}({\bf r},z) + \frac{1}{1-2\sigma}\nabla \nabla \cdot \vct{u}({\bf r},z)  = \frac{-2\rho_s (1+\sigma)}{E} \vct{F}({\bf r}, z), 
\end{eqnarray}
which, since it is linear, can be solved by the {\sl Ansatz}
\begin{eqnarray}
u_i ({\bf r},z) = \int_V {\cal G}_{ik} ({\bf r} - {\bf r}', z-z') F_k ({\bf r}',z') ~d^2{\bf r} dz,  
\label{fund-navier}
\end{eqnarray}
where the {\sl }elastic Green's dyadic has to satisfy
\begin{eqnarray}
\left(\bnabla^2 \delta_{il} + \frac {1}{1-2
\sigma} \bnabla_i \bnabla_{l} \right){\cal G}_{lk} ({\bf r} - {\bf r}', z-z')
= - \frac {2 \rho_s (1+\sigma) }{E} \delta_{ik} \delta^{2}({\bf r} - {\bf r}') \delta(z-z').
\end{eqnarray}

Inserting the Green's dyadic into the elastic free energy we remain with
\begin{eqnarray}
{\cal W} = - {\textstyle\frac12} \rho_s^2~\int_S\int_z  {\cal G}_{ik} ({\bf r} - {\bf r}', z-z')~ F_{i}({\bf r}, z)~F_k ({\bf r}',z') ~d^2{\bf r}' dz'.
\label{appenerg}
\end{eqnarray}
If the external force is given by two point like sources acting at the upper surface, $z=0$, in the $z$ direction,  then with Eq.~\eqref{gfdhjksa1}
\begin{equation}
    \vct{F}({\bf r},z)= \left(~0~, ~0~, ~\delta(z)~\left(P_1 \delta^2({\bf r}-{\bf r}_1) + P_2 \delta^2({\bf r} -{\bf r}_2)\right)\right),
    \label{gfdhjksa2}
\end{equation}
we obtain from Eq.~\eqref{appenerg} the interaction free energy between these two point like sources in the form
\begin{eqnarray}
 {\cal W} = - {\textstyle\frac12} \rho_s^2 ~{\cal G}_{zz} ({\bf r}_1 - {\bf r}_2, z=z'=0)~ P_1 ~P_2,
\end{eqnarray}
where we ignored the two formally infinite self-energy terms that do not modify the interaction between the sources. This is exactly Eq.~\eqref{elastic_interact} in the main text. The whole derivation very much follows the standard approach in electrostatics. 


\section{Elastic displacement and stress tensor}
\label{appendix:a}

In this appendix, we provide explicit expressions for the elastic displacement and stress tensor with the Galerkin-Love {\sl Ansatz}~\cite{gal36}, The Galerkin vector and the Love solution are introduced as follows
\begin{eqnarray}
\mathbf{u}({\bf r},z) = \nabla^2\mathbf{g}({\bf r},z) - \frac{1}{2(1 - \sigma)}\nabla (\nabla \cdot \mathbf{g}({\bf r},z)) \qquad {\rm with} \qquad {\bf g}= (0, 0, Z({\bf r},z)).
\label{solveg1}
\end{eqnarray}

From here, the elastic displacement components are expressed as
\begin{subequations}
\begin{eqnarray}
u_{x} &=& - \frac{1}{2(1 - \sigma)}\frac{\partial}{\partial x} \frac{\partial Z(r,z) }{\partial z}  \qquad u_{y} = u_{x}(x\longrightarrow y)
\label{displacement_xy}
\end{eqnarray}
and
\begin{eqnarray}
u_{z} &=& \nabla^{2} Z(r,z) - \frac{1}{2(1 - \sigma)}\frac{\partial^{2}Z(r,z)}{\partial z^{2}}.
\label{displacement_z}
\end{eqnarray}
\end{subequations}

From standard definitions the stress tensor the off diagonal components $p_{xy}, p_{xz}, p_{yz}$ then follow as 
\begin{eqnarray}
p_{xy} = \frac{E}{1+\sigma}u_{xy} = \frac{-E}{4(1 - \sigma^2)}\frac{\partial^2}{\partial x \partial y}\frac{\partial Z(r,z)}{\partial z}
\label{p_xy}
\end{eqnarray}

\begin{eqnarray}
p_{xz} = \frac{E}{1 + \sigma}u_{xz} = \frac{E}{2(1 - \sigma^2)}\left((1 - \sigma)\nabla^2 - \frac{\partial^2 }{\partial z^2}\right)\frac{\partial Z(r,z)}{\partial x}  \qquad p_{yz} = p_{xz}(x \longrightarrow y)
\label{p_xz}
\end{eqnarray}
while the diagonal components read
\begin{eqnarray}
p_{xx} = \frac{E}{2(1 - \sigma^2)}\left(\sigma \nabla^2 - \frac{\partial^2}{\partial x^2} \right)\frac{\partial Z(r,z)}{\partial z} \qquad p_{yy}= p_{xx}(x \longrightarrow y)
\label{p_xx}
\end{eqnarray}
\begin{eqnarray}
p_{zz} = \frac{E}{2(1 - \sigma^2)}\left((2 - \sigma) \nabla^2 - \frac{\partial^2}{\partial z^2} \right)\frac{\partial Z(r,z)}{\partial z},
\label{p_zz}
\end{eqnarray}
where the arrows indicate the two quantities have the same form except for the substitution indicated by the arrow.

\section{The Love potential}
\label{appendix:b}

Consistent with the cylindrical symmetry of the problem and isotropy in the $(x, y)$ plane, one can introduce the Fourier-Bessel transform $Z(Q =\vert{\bf Q}\vert, z)$, see Eq.~\eqref{ft_stress_function}, so that the homogeneous Love equation, Eq.~\eqref{def_z}, is solved by 
\begin{eqnarray}
Z(Q,z) = \Big((A + BQz)e^{-Qz} + (C + DQz)e^{Qz}\Big),
\label{1four_coefficients}
\end{eqnarray}
where the integration constants $A$, $B$, $C$ and $D$ are determined from the boundary conditions, which we consider next. 

 At the upper surface, $z = 0$, the boundary condition Eqs. ~\eqref{pz_0} and \eqref{pxpy_01} specify the components of the stress tensor. As the force is applied only in the $z-$ direction the boundary condition at that surface reads,
\begin{eqnarray}
p_{xz}(Q, z = 0) = p_{yz}(Q, z = 0) = 0 \qquad {\rm and} \qquad 
p_{zz}(Q, z = 0) = {P}.
\label{pzz}
\end{eqnarray}
expressed in terms of the Fourier-Bessel components of the stress tensor.  The first two identities of Eq.~\eqref{pzz} are satisfied by 
\begin{eqnarray}
\left((1 - \sigma)(\frac{\partial^2}{\partial z^2} - Q^2) - \frac{\partial^2}{\partial z^2}\right)Z(Q,z=0) = 0,
\label{z00}
\end{eqnarray}
while the last one implies
\begin{eqnarray}
\frac{\partial}{\partial z}\left((2 - \sigma)(\frac{\partial^2}{\partial z^2} - Q^2) - \frac{\partial^2}{\partial z^{2}} \right)Z(Q,z=0) = \frac{2 \rho_s (1-\sigma^2)~P}{E}.
\label{p/e}
\end{eqnarray}
Furthermore, Eq.~\eqref{z00} yields
\begin{eqnarray}
-(A - 2\sigma B) - C - 2\sigma D = 0
\label{z=0}
\end{eqnarray}
and Eq.~\eqref{p/e},

\begin{eqnarray}
A - 2\sigma B - C - 2\sigma D + B + D = \frac{2 \rho_s (1 - \sigma^2)~P}{EQ^3}.
\label{z=02}
\end{eqnarray}

At the lower surface, $z = -h$, the boundary condition is simply that all the components of the deformation vector vanish at this plane. 
\begin{equation}
    u_x(Q, z = -h) = 0,~u_y(Q, z = -h) = 0,~u_z(Q, z = -h) = 0,
\end{equation}
expressed in terms of the Fourier-Bessel components of the displacement vector, or explicitly
\begin{eqnarray}
 \frac{\partial Z(Q,z = -h)}{\partial z} = 0
 \end{eqnarray}
 and
 \begin{eqnarray}
 \Big(\frac{1 - 2\sigma}{2(1-\sigma)} \frac{\partial^{2}}{\partial z^{2}} - Q^2\Big) Z(Q,z = -h) = 0,
\label{disp_z}
\end{eqnarray}
which implies, Eq.~\eqref{z00}, that 
\begin{eqnarray}
(-A + B + BQh)Qe^{Qh} + (C + D -DQh)Qe^{-Qh} = 0
\label{z=-h1}
\end{eqnarray}
while Eq.~\eqref{displacement_z} yields
\begin{eqnarray}
-Ae^{Qh} + (-2 + 4\sigma + Qh)Be^{Qh} - Ce^{-Qh} + (2-4\sigma + Qh)De^{-Qh} =0.
\label{z=-h2}
\end{eqnarray}
Both boundary conditions, Eqs.~\eqref{z=0}, \eqref{z=02}, \eqref{z=-h1} and \eqref{z=-h2}, can be cast as a system of linear equations with undetermined coefficients
\begin{eqnarray}
\begin{vmatrix}
-e^{Qh} && (-2 + 4\sigma  + Qh)e^{Qh} && -e^{-Qh} && (2 -4\sigma + Qh)e^{-Qh} \\
-e^{Qh}  && (1 + Qh)e^{Qh} && e^{-Qh} && (1 - Qh)e^{-Qh} \\
1 && 1- 2\sigma && -1 && 1 - 2\sigma \\
1 && -2\sigma && 1 && 2\sigma
\end{vmatrix}
\begin{vmatrix}
A \\
B \\
C \\
D
\end{vmatrix}
= 
\begin{vmatrix}
0 \\
0 \\
\frac{2 \rho_s (1-\sigma^2)}{EQ^3}\\
0
\end{vmatrix}.
\label{hopf}
\end{eqnarray}

We assume furthermore that the determinant of this system 
\begin{eqnarray}
   {\rm Det} = (3- 4\sigma) + \left((3 - 4\sigma)^2 + 1 + (2hQ)^2\right) e^{2Qh} + (3- 4\sigma) e^{4Qh}
\end{eqnarray}
is non-vanishing and an explicit solution can therefore be obtained.


\section{Solution of the boundary conditions and the Boussinesq limit}

The values of coefficients $A$, $B$, $C$ and $D$ in Eq.~ \eqref{hopf} are then found to be as follows,
\begin{eqnarray}
A = \frac{2 \rho_s P(1-\sigma^2)(6\sigma -8\sigma^2 +4e^{2Qh} + 2e^{2Qh}h^2Q^2 -10e^{2Qh}\sigma + 4e^{2Qh}hQ\sigma + 8e^{2Qh}\sigma^2)}{EQ^3 ~{\rm Det}}
\end{eqnarray}
\begin{eqnarray}
B = \frac{2 \rho_s P(1-\sigma^2)\bigg(3 -4\sigma +e^{2Qh} + 2e^{2Qh}Qh\bigg)}{E Q^3~ {\rm Det}}
\end{eqnarray}
\begin{eqnarray}
C = \frac{2 \rho_s P(1-\sigma^2)(-4e^{2Qh} - 2e^{2Qh}h^2Q^2 + 10e^{2Qh}\sigma - 6e^{4Qh}\sigma + 4e^{2Qh}hQ\sigma - 8e^{2Qh}\sigma^2 + 8e^{4Qh}\sigma^2)}{EQ^3 ~{\rm Det}}
\end{eqnarray}
\begin{eqnarray}
D = \frac{2 \rho_s P(1-\sigma^2)(e^{2Qh} + 3e^{4Qh} - 2e^{2Qh}hQ - 4e^{4Qh}\sigma)}{E Q^3~ {\rm Det}}.
\label{BCs}
\end{eqnarray}
The Love potential can then be evaluated from Eq.~\eqref{1four_coefficients} as
\begin{eqnarray}
Z(r,z) = \frac{1}{2\pi}\int_{0}^{\infty}QdQJ_{0}(Q r)\Big((A + BQz)e^{-Qz} + (C + DQz)e^{Qz}\Big).
\label{ft_stress_function2}
\end{eqnarray}
This is the solution of the Boussinesq problem for a finite thickness elastic matrix with a point force acting perpendicular to the upper surface. Note that in the limit $h\longrightarrow \infty$ only the coefficients $A$ and $B$ remain finite, so that the original Boussinesq solution is recovered
\begin{eqnarray}
\lim_{h \longrightarrow \infty}Z(r,z) =  \frac{2  \rho_s P (1-\sigma^2)}{2\pi E}\int_{0}^{\infty}Q^{-2}dQJ_{0}(Q r) (2 \sigma + Qz)e^{-Qz}.
\label{Boussinesq}
\end{eqnarray}
Expressing the displacement field with the Love potential, Eqs.~ \eqref{displacement_z}, we get
\begin{eqnarray}
u_z(r, z) = \frac{1}{2\pi}\int_{0}^{\infty}QdQJ_{0}(Q r) u_z(Q, z) = \frac{1}{2\pi}\int_{0}^{\infty}QdQJ_{0}(Q r)\left(Q^2 Z(Q,z) - \frac{1 - 2\sigma}{2(1 - \sigma)}\frac{\partial^{2}Z(Q,z)}{\partial z^{2}}\right),
\label{1FBGreens1}
\end{eqnarray}
and analogously for $u_{x,y}$. Inserting the solution Eq.~\eqref{ft_stress_function2} we are left with
\begin{eqnarray}
u_z(Q, z) = \frac{ \rho_s P (1+\sigma)}{2 E Q (\sigma -1)}{\left(e^{Q z} (B (Q z +4 \sigma -2)-A)+(D (Q z -4 \sigma +2)-C)e^{-Qz}\right) },
\label{FBGreens2}
\end{eqnarray}
and again analogously for $u_{x,y}$ with Eq.~\eqref{displacement_z}. From here we get back the solution $u_{z}(r, z)$ to the original Boussinesq problem of a point force at the origin with $h \longrightarrow \infty$,
\begin{eqnarray}
\lim_{h \longrightarrow \infty}u_z(r, z) = 
\frac{ \rho_s P (1+\sigma)}{2\pi E}\int_{0}^{\infty}  dQJ_{0}(Q r)\left( Q z + 2(1-\sigma)\right) e^{-Q z},
\label{FBGreens3}
\end{eqnarray}
wherefrom it follows
\begin{eqnarray}
u_{z}(r, z) &=& \frac{ \rho_s (1 + \sigma)P}{2\pi E}\left(\frac{z^{2}}{(r^{2} + z^{2})^{3/2}} + \frac{2(1 - \sigma)}{(r^{2} + z^{2})^{1/2}} \right) = {\cal G}_{zz}(r, z) ~\rho_s P 
\label{eq:38}
\end{eqnarray}
and analogously for $u_{(x,y)}(r, z)$
\begin{eqnarray}
 u_{(x,y)}(r, z) &=& \frac{ \rho_s (1 + \sigma)P}{2\pi E}\left(\frac{-xz}{(r^{2} + z^{2})^{3/2}} + \frac{(1 - 2\sigma)x}{(r^{2} + z^{2})^{1/2}((r^{2} + z^{2})^{1/2} + \vert z\vert)} \right) = {\cal G}_{(x,y)z}(r, z) ~\rho_s P ,
\end{eqnarray}
where we wrote the solution with the components of the Green's dyadic, see Eq.~\eqref{fund-navier}. It is clear form this solution that the elastic deformation diverges at the origin for $r = 0, z = 0$. This is of course an artifact, as the Boussinesq calculation works only for linear elasticity valid for small deformations, while the deformation close to the application point of the external force can be large. This spurious divergence can be eliminated by assuming a finite region of the application of the force and truncating the solution within this region. 

In Fig.~\eqref{uzprofile} we display the contours of the $z$ component of the deformation vector, $u_z(r, z)$. Clearly the effect of the boundary condition at the bottom surface, {\sl viz.}, $u_{x,y,z}(r, z = h) = 0$ is the contraction of the profile in the $z$ direction, see the contour plot in Fig.~\eqref{uzprofile}. This affects also the interaction potential between the point sources or consequently the extended rods.

\section{Deformation of the top surface}
What is an observable is the deformation vector at the top surface, whose $z$ component we consider next. The deformation $u_z$ of the top surface is now obtained from the Green's function ${\cal G}_{zz}$ defined in Eq.~\eqref{Greens}. This Green's function at the same time defines also the interaction between two point-like surface inclusions. In the Boussinesq limit of an infinitely thick substrate, Eq.~\eqref{eq:38}, we can extract
\begin{eqnarray}
{\cal G}_{zz}(r, z=0)& =& 
 \frac{2 (1 - \sigma^2)}{2\pi E}\frac{1}{r},
\label{2GreenBouss}
\end{eqnarray}
which is of an electrostatic form. 
We now proceed to the general case of finite thickness of the substrate where we get
\begin{eqnarray}\nonumber
{\cal G}_{zz}(r, z=0) &=& 
\frac{1}{2\pi}\int_0^\infty Q ~{\cal G}_{zz}(Q, z = 0) J_0(Q r)dQ
\label{hatzp}
\end{eqnarray}
with
\begin{eqnarray}
    {\cal G}_{zz}(Q, z = 0) =  \frac{2   (1-\sigma^2)}{E~Q} \left(\frac{-(3-4\sigma) - 4Qh e^{2Qh} + (3-4\sigma)e^{4Qh}}{(3 - 4\sigma) + \left((3 - 4\sigma)^2 + 1 + (2Qh)^2\right) e^{2Qh} + (3 - 4\sigma) e^{4Qh}}\right) = \frac{2 (1-\sigma^2)}{E} \frac{\tilde{\cal G}_{zz}(Qh)}{Q}. \nonumber\\
    ~
    \label{gfhajl}
\end{eqnarray}
From here we obtain
\begin{eqnarray}
{\cal G}_{zz}(r, z=0) &=& 
\frac{2 (1-\sigma^2)}{2\pi E} \frac{1}{r} \int_0^\infty \tilde{\cal G}_{zz}\left(u\frac{h}{r}\right) J_0(u)du = \frac{2 (1-\sigma^2)}{2\pi E} \frac{1}{r} ~{\cal H}_{\sigma}\left(\frac{h}{r}\right),
\label{hatzp1}
\end{eqnarray}
where ${\cal H}_{\sigma}\left(x\right)$ fully quantifies the dependence on the thickness of the substrate on top of the Coulombic dependence for an infinitely thick substrate, Eq.~\eqref{2GreenBouss}.  The dependence ${\cal H}_{\sigma}\left(x\right)$ is shown on Fig.~\eqref{FunH}. From this functional form it is clear that for large thickness of the substrate
\begin{eqnarray}
\lim_{x\longrightarrow \infty} {\cal H}_{\sigma}\left(x\right) = 1,
\end{eqnarray}
and we are back to the  Coulombic dependence for the pure Boussinesq case, Eq.~\eqref{2GreenBouss}. On the other hand for $x \leq 0.65$ ${\cal H}_{\sigma}\left(x\right)$ changes sign and approaches zero. In the case of two surface point sources at ${\bf r}_1$ and ${\bf r}_2$, the elastic interaction between them is proportional to ${\cal G}_{zz}(\vert{\bf r}_1-{\bf r}_2\vert,z=0)$, see Eq.~\eqref{elastic_interact}. We thus conclude that the interaction between two point surface sources is almost Coulombic for small separations but is then effectively cut off for $\rho \geq h$. It looks as if the interactions are ``screened". This effect is due purely to the finite thickness of the substrate. 


\section{Calculation of  $a^{(2)}$ }
\label{AppF}

We start from Eq.~\eqref{bdfghjq} and Eq.~\eqref{hatzp20}, where we derived explicitly
\begin{eqnarray}
a^{(2)} &=& \frac{8}{9} \int_0^{\infty} r dr V^{(2)}(r)
\label{2bdfghjq11}
\end{eqnarray}
with
\begin{eqnarray}
V^{(2)}(r) &=&  -\frac{2 (1-\sigma^2)}{2\pi E} \frac{\rho_s^2 ~(p L)^2~L^4}{2304} 
  \frac{1}{r^5} {\cal F}_{\sigma}\left(\frac{h}{r}\right). 
\label{hatzp21}
\end{eqnarray}
This furthermore implies that 

\begin{eqnarray}
a^{(2)} &=& -\frac{8}{9} \frac{(1-\sigma^2)}{2304~\pi E} \rho_s^2 ~(p L)^2~L \left(\frac{L}{h}\right)^3 \int_0^{\frac{h}{L}} x^2 dx  ~{\cal F}_{\sigma}(x) = \nonumber\\
&& ~-\frac{8}{9} \frac{(1-\sigma^2)}{2304~\pi E} \rho_s^2 ~(p L)^2~L ~\tilde{a}^{(2)}
\label{3bdfghjq13}
\end{eqnarray}
where $\tilde{a}^{(2)}$ is the dimensionless orientational virial coefficient and ${\cal F}_{\sigma}(x)$ has been defined in Eq.~\eqref{hatzp20}, while the infinite upper bound has been substituted by $h/L$, 
This upper bound is necessary since otherwise the integral is divergent. 
The asymptotic limit of the orientational virial coefficient for large values of the argument is
\begin{eqnarray}
\lim_{\frac{h}{a}\longrightarrow \infty} \tilde{a}^{(2)} = 30,
\label{lgval1}
\end{eqnarray}
which follows from the definition of the integrand ${\cal F}_{\sigma}(x)$ and its properties in the asymptotic limit. As the integrand is also a very slowly limiting function, its limit is also reached only asymptotically. The above results are used in the analysis of the orientational virial coefficient in the main text.

\end{document}